\documentclass[aps,prl,reprint,amsmath,amssymb,superscriptaddress,citeautoscript,floatfix]{revtex4-1}
\usepackage{graphicx}
\usepackage{color}
\renewcommand{\vec}[1]{\mathbf{#1}}
\newcommand{\isoavg}[1]{\langle {#1} \rangle_{\text{iso}}}

\newcommand{\supcite}[1]{\cite{#1-sup}}
\newcommand{\onlinesupcite}[1]{\onlinecite{#1-sup}}

\begin{document}
\title{Correlation of local order with particle mobility in supercooled liquids is highly system dependent}

\author{Glen M. Hocky}
\email{gmh2123@columbia.edu}
\affiliation{Department of Chemistry, Columbia University, 3000 Broadway, New York, New York 10027, USA}

\author{Daniele Coslovich}
\email{daniele.coslovich@univ-montp2.fr}
\author{Atsushi Ikeda}
\email{atsushi.ikeda@univ-montp2.fr}
\affiliation{CNRS, Laboratoire Charles Coulomb UMR 5221, Montpellier, France}
\affiliation{Universit\'{e} Montpellier 2, Laboratoire Charles Coulomb UMR 5221, Montpellier, France}

\author{David R. Reichman}
\email{drr2103@columbia.edu}
\affiliation{Department of Chemistry, Columbia University, 3000 Broadway, New York, New York 10027, USA}

\date{\today}

\begin{abstract}
We investigate the connection between local structure and dynamical heterogeneity in supercooled liquids.
Through the study of four different models we show that the correlation between a particle's mobility and the degree of local order in nearby regions is highly system dependent.
Our results suggest that the correlation between local structure and dynamics is weak or absent in systems that conform well to the mean-field picture of glassy dynamics and strong in those that deviate from this paradigm.
Finally, we investigate the role of order-agnostic point-to-set correlations and reveal that they provide similar information content to local structure measures, at least in the system where local order is most pronounced.
\end{abstract}

\maketitle


Supercooled liquids display markedly heterogeneous dynamics, despite possessing structural properties that appear nearly unchanged from those of normal liquids from which they are prepared~\cite{Ediger-ARPC2000}.
While there has been intense focus on understanding dynamical heterogeneity in a wide variety of systems, the structural origin of this phenomenon is not well understood~\cite{Berthier-Book2011,Berthier-RMP2011}.
Simulations of model supercooled liquids are useful for understanding the connections between structure and dynamics because particle locations may be followed precisely for all times.
Nonetheless, new theoretical tools are needed to filter out extraneous detail from the key structural and dynamical fluctuations in glassy systems.

One particularly useful simulation-based tool for quantifying the influence of structure on dynamics is the isoconfigurational ensemble, where a large number of molecular dynamics (MD) simulations are initiated from the same starting configuration with momenta sampled randomly from a Boltzmann distribution~\cite{WidmerCooper-PRL2004,WidmerCooper-JPCM2005}.
Under glassy conditions, spatial heterogeneities are immediately evident in the isoconfigurational displacement (or propensity) field.
A reasonable hypothesis is that particles with low propensity have a larger measure of local structural stability.
Surprisingly, however, simple structural quantities, such as free volume and local potential energy, show little correlation with the heterogeneity of the propensity field~\cite{WidmerCooper-JNCS2006}.
In some models, localized soft modes~\cite{WidmerCooper-NatPhys2008,WidmerCooper-JCP2009,Candelier-PRL2010} or unstable modes~\cite{Coslovich_Pastore_2006} appear to correlate strongly with propensity, 
but the degree of universality of this connection has not been thoroughly investigated.

Recently, focus has turned to the study of specific structural motifs and their putative connection with the dynamics of supercooled liquids. The notion that the frustration of local order incommensurate with bulk crystalline periodicity may be related to glass formation is an old one~\cite{Frank-PRS1952,Steinhardt-PRL1981,Nelson-2002Defects,Kivelson-PhysicaA1995}.
New evidence for the growth of domains associated with local packing motifs has been presented for several simple~\cite{Coslovich-JCP2007} and realistic model systems~\cite{Ding_Cheng_Sheng_Ma_2012}, where particles tend to be found in certain ``locally preferred structures'' (LPS) with increased supercooling. As a general rule, more fragile systems display a more rapid increase in LPS concentration and domain extent~\cite{Coslovich-JCP2007,Ding_Cheng_Sheng_Ma_2012}. In some systems, a correlation between the size and location of LPS and slow dynamics has been observed~\cite{Tanaka_Kawasaki_Shintani_Watanabe_2010,Malins-JCP2013}, although the quantitative meaning of the correlations observed remains, in a statistical sense, obscure.

Point-to-set (PTS) correlations have emerged as an alternative quantifiable metric of amorphous ordering. PTS correlations measure the decrease of configurational entropy imposed by the presence of particles pinned in an equilibrium configuration~\cite{Bouchaud-JCP2004,Berthier-static-PRE2012}.
The length scale associated with PTS correlations has been demonstrated to grow upon increased supercooling in several systems~\cite{Biroli-NatPhys2008,Hocky-PRL2012,Berthier-static-PRE2012},
although its variation is rather modest over the dynamical range currently accessible in simulations~\cite{Charbonneau_Charbonneau_Tarjus_2012}, 
Nonetheless, several observations indicate that the growing PTS length scale should ultimately drive the dramatic increase in relaxation times in supercooled liquids~\cite{Karmakar-PNAS2009,Biroli-PRL2013,Karmakar-ARCMP2014}. 
It should be noted that PTS correlations, as well as other recently proposed measures of static correlations \cite{Ronhovde-EPJE2011,Sausset-PRL2011}, are ``order agnostic''~\cite{Charbonneau_Charbonneau_Tarjus_2012} and therefore their growth does not necessarily connect to the emergence of specific local structures, such as those identified in the LPS studies. 

\begin{figure}[t]
\centering
\includegraphics[width=3.2in]{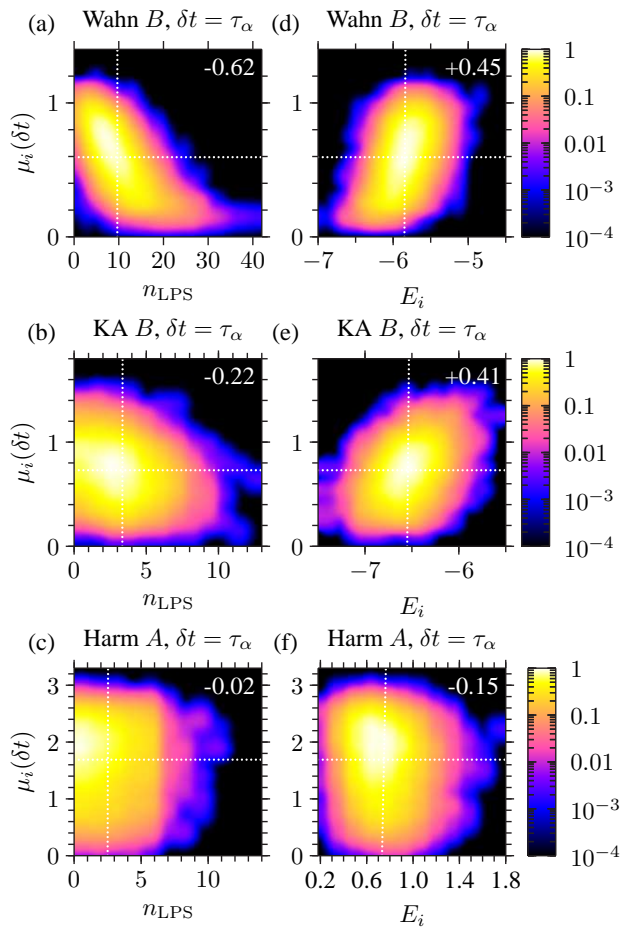}
\caption{Interpolated histograms of particle mobility. Numbers at top-right indicate Spearman rank correlation coefficients $K$~\cite{Sammut-Encyclopedia2011}. The first column shows the correlation between mobility and $n_{\rm LPS}$ with the LPS defined in the text. The second column shows correlation with $E_i$, the sum of a particle and its neighbors' pair energies. White dotted lines show the average value of the quantity on the horizontal and vertical axes.}
\label{fig:lpshist}
\end{figure}

In this Letter we quantify the correlation between between static structure and dynamical heterogeneity in supercooled liquids in a statistically precise sense and within a coherent  simulation framework. 
We demonstrate that seemingly similar systems may differ dramatically with respect to the degree to which specific local structural motifs correlate with dynamics. 
Our results indicate that scenarios connecting LPS cluster formation and glassy behavior~\cite{Tanaka_2012,Langer_2013} cannot be generically correct.
The observed model dependence suggests instead that local structural quantities play a key role in systems with large deviations from mean-field glassy behavior.
Lastly, we show that a connection exists between growing PTS correlations and LPS in systems where LPS are strongly predictive of dynamical heterogeneity. 

The first two models we will study are binary Lennard-Jones mixtures, namely the Kob-Andersen (KA) system ~\cite{Kob-PRL1994} and the Wahnstr\"{o}m (Wahn) system~\cite{Wahnstrom-PRA1991}. The definition of these models and their LPS statistics have been extensively detailed in Ref.~\onlinecite{Coslovich-PRE2011}.
The KA system is an 80:20 mixture while the Wahn system is equimolar.
As a third system, we study a binary mixture of harmonic spheres (Harm) at a density such that that jamming is approached by lowering temperature near to zero ($\rho=0.675$)~\cite{Kob-NatPhys2011}. In all cases, one species is smaller ($B$ for KA and Wahn, $A$ for Harm) and is intrinsically more mobile. Results for the small particles will be reported in the main text and for the large particles in the Supplemental Information (SI)\footnote{
See Supplemental Material [url], which includes Refs. \cite{Kim_Saito_2013,OHern_Langer_Liu_Nagel_2002,Stillinger_1976-sup,Lang_Likos_Watzlawek_Lowen_2000-sup,tanemura_geometrical_1977,Biroli-JPCM2007,
Eaves-PNAS2009-sup,Charbonneau-PRE2010-sup,Charbonneau-JCP2013-sup}.}.
In the following, we will discuss all quantities using standard reduced units. For all three systems, we study systems with $N=1000$. Further simulation details and a description of the LPS in each system can be found in the SI. 

\begin{figure*}[t]
\centering
\includegraphics[scale=0.95]{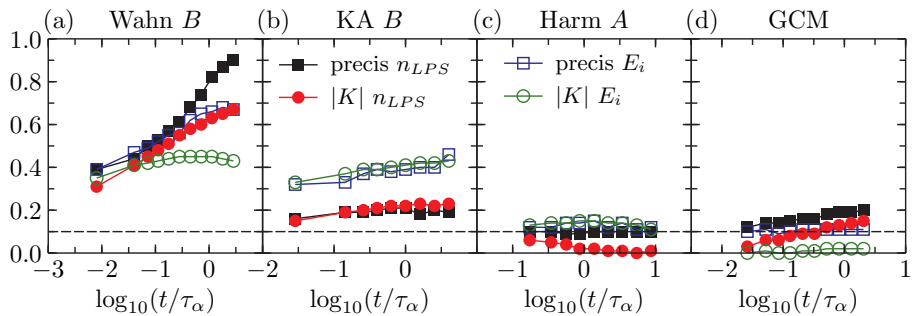}
\caption{Time dependence of predictability metrics and correlation coefficients as a function of time for four models. Closed symbols correspond to data comparing mobility with $n_{\rm LPS}$ as in the left-hand column of Fig.~\ref{fig:lpshist} and open symbol comparing local energy $E_i$ (defined in text) to mobility, as in the right-hand column of that figure. $K$ is the Spearman rank correlation coefficient~\cite{Sammut-Encyclopedia2011}. Precision is defined as the percentage of particles in the the top (bottom) $2\%$ of $n_{\rm LPS}$ ($E_i$) which are in the bottom $10\%$ of $\mu_i$. Horizontal dashed lines show the result for the precision metrics that would result from random classification of particles as slow. }
\label{fig:predict}
\end{figure*}

While all of the above models are simple binary mixtures with short-ranged interaction potentials, their local structures differ significantly.
For each model, we identify particles participating in LPS through a Voronoi analysis (see SI).
These LPS correspond to icosahedra, bicapped square anti-prisms, and distorted icosahedral structures in Wahn, KA, and Harm respectively.
The relative abundance of these LPS at low temperatures is model dependent: it is fairly significant in the Wahn model (about 10\% of the particles are at the center of a LPS) and weaker in the other models. 
In the KA model this is due to the fact that twisted bicapped prisms are mostly centered around the small particles, which constitute the minority species.

In order to investigate the connection between the LPS in each system and dynamical behavior, we perform simulations in the isoconfigurational ensemble at supercooled temperatures, $T=0.588$ for the Wahn, $T=0.45$ for the KA and  $T=5.5$ for the Harm.  
These temperatures correspond roughly to the same degree of supercooling, as measured by the relative distance (about 3--6\%) from their fitted Mode-Coupling temperatures, $T_c$ (see SI). 

We select 40 (20) equilibrated configurations for the KA and Wahn (Harm) systems and perform the Voronoi analysis as discussed above. For each configuration, we performed 200 (100) $NVT$ simulations in the isoconfigurational ensemble.
From the simulations starting from each configuration, we compute the particle mobilities $\mu_i(t) \equiv \isoavg{| \vec{r}_i(t) - \vec{r}_i(0) |} \equiv \isoavg{ |\delta \vec{r}_i(t) |}$. 
To quantify the number of LPS associated with a given particle, we count the number of structures deemed locally preferred in a spherical region of radius $l$ around each particle ($n_{\rm LPS}$).
All results reported here are nearly insensitive to this $l$ value in the range we have investigated $1.5\leq l \leq3.0$, and we chose to report results only for $l=2.5$.


In Figs.~\ref{fig:lpshist}(a)-(c) we show the combined probability distribution of $\mu_i$ and $n_{\rm LPS}$ for the 3 systems introduced above. We quantify correlation by using the Spearman rank correlation coefficient $K$~\cite{Sammut-Encyclopedia2011}, which has been used previously in a similar context~\cite{WidmerCooper-JNCS2006}. $K$ is $1$ if two quantities are related by a monotonically increasing function and $-1$ if by a decreasing one. $K$ values are shown in the top-right corner of each histogram. We see visually and quantitatively that the correlation is much stronger in the Wahn system than in the KA and Harm systems. For comparison, the probability distributions for correlation between mobility and local energy ($E_i=(e_i+\sum_{j\in\text{neigh}(i)} e_j)/(1+|{\rm neigh}(i)|$) , where $\rm |neigh(i)|$ is the number of neighbors in the Voronoi structure around particle $i$) are shown in Figs.~\ref{fig:lpshist}(d)-(f). The correlation is fairly significant in the two Lennard-Jones mixtures and much weaker in the Harm system. 

Inspection of Fig.~\ref{fig:lpshist}(a) reveals a long tail in the histogram of $n_{\rm LPS}$ values.
From these data, we can predict that a particle in a domain rich in icosahedral structures will be very immobile. Looking at the data at $n_{\rm LPS}=0$ instead, we see that such particles will have higher than average mobility. However, slow and fast particles have a wide range of $n_{\rm LPS}$ values. In Fig.~\ref{fig:predict} we show the level of ``precision''~\cite{Sammut-Encyclopedia2011} in predicting which particles are slow based on $n_{\rm LPS}$ and $E_i$ for the models studied,  as well as the $K$ values.
Here, precision is defined as the percentage of particles in the top (bottom) 2\% of $n_{\rm LPS}$ ($E_i$) which are also in the bottom 10\% of $\mu_i(\delta t)$. All trends discussed are insensitive to the particular percentiles chosen for this definition of precision (see SI for further details, including a discussion of sample-to-sample fluctuations in these quantities).

When viewed from this statistical perspective, several striking features are observed. For the Wahn system, LPS are highly predictive of slow dynamics. In particular, using our definition of precision one may ``predict'' the location of slow particles with near perfect accuracy up to $\tau_\alpha$, and such a correlation continues to grow to the longest times we investigated. In the KA system, local energy is  more predictive of slow dynamics than LPS locations, and correlation for both local energy and $n_{\rm LPS}$ in KA and Harm are far lower than in the Wahn system.

We may thus conclude that the correlation between dynamics and local structural metrics such as $n_{\rm LPS}$ is highly system dependent. What may be taken from this dramatic degree of variability?
Among the three models studied, the Wahn system shows the largest departures from mean-field behavior. Namely, Wahn exhibits large violation of the Stokes-Einstein relation, sizable deviations from time-temperature superposition, and large inconsistencies between fitted Mode-Coupling exponents~\cite{Schrøder_Sastry_Dyre_Glotzer_2000} (see SI). From this perspective, the KA system shows moderate deviations from mean-field behavior. 
This leads us to consider whether the correlation between local structural order and slow dynamics might be connected to how much a model system conforms to the mean-field paradignm. While the Harm system does not uniformly display mean-field behavior, results from finite size studies~\cite{Berthier-finite-PRE2012} and the existence of a non-monotonic dynamical length scale~\cite{Kob-NatPhys2011} suggest that its behavior is at least partially harmonious with mean-field theory. This leads us to posit a connection between a high degree of local structure-dynamics correlation and strong spatial fluctuations which are manifest in systems that deviate from mean-field behavior.

To better test this notion, we study a fourth system, 
the high-density ($\rho=2.0$) Gaussian Core Model (GCM).
The GCM is a single-component fluid with Gaussian repulsions~\cite{Ikeda-PRL2011,Ikeda-JCP2011}, which has all the hallmarks of glassy behavior while matching mean-field predictions of dynamical exponents, strongly suppressed non-Gaussian fluctuations and minimal Stokes-Einstein violation~\cite{Ikeda-JCP2011}.
This mean-field behavior seems to arise naturally from the long ranged and ultra-soft interaction potential (see discussion in SI).

In Fig.~\ref{fig:predict}(d) we show results for $N=3456$ Gaussian core particles at $T=3.2$ with 100 isoconfigurational runs initiated from 20 independent configurations. We note that this temperature is slightly higher, relative to $T_c=2.7$, than the one used in the other models, but corresponds instead to the same relative increase in relaxation time as observed for the Wahn system (see SI)
We found that distorted crystal-like structures constitute the LPS of the model (the underlying stable crystal at the studied density is BCC). In agreement with our expectations, the correlation between $n_{\rm LPS}$ and dynamics in the GCM system is very low, just as in the Harm system, and only marginally improves as $t$ increases
\footnote{It may appear that the correlation between $n_{\rm LPS}$ and dynamics in both the KA and GCM system are superficially similar. However, the LPS in the GCM system are simple crystalline motifs that exist because of the difficulty of avoiding such particle arrangements in a monatomic system. In this sense, we view the correlation of {\em both} $n_{\rm LPS}$ and $E_i$ as significantly weaker in the GCM system when compared to KA.}. 

It would be natural to speculate that in systems such as the Harm and GCM models, there simply exists {\em no} connection between structure and dynamics. However, this statement is incorrect. We have used the $R_4$-ratio analysis of Berthier and Jack~\cite{Berthier-PRE2007} to quantify the structural component of the dynamic fluctuations. As detailed in the SI, we found that all four systems analyzed in Fig.~\ref{fig:predict} show a marked correlation between structure and dynamics, despite the fact that no {\em specific} structural motif connects to dynamics in the more mean-field like models. These striking results will be a subject for future investigations. Here we just point out an interesting analogy with the behavior of mean-field $p$-spin models~\cite{Franz-EPJE2011}, which \textit{do} display large values of $R_4$ close to the dynamic transition.

One may take the inability of specific structural metrics, such as LPS determined from Voronoi analysis, to correlate universally with dynamics in supercooled liquids as an indication that a more general form of growing amorphous order must be implicated. In the remaining of this work we focus on structural correlations embodied in point-to-set and related length scales~\cite{Biroli-NatPhys2008,Hocky-PRL2012,Biroli-PRL2013,Karmakar-ARCMP2014}. In order to show that this type of order may subsume specific structural metrics, we investigate the connection between local order as measured by $n_{\rm LPS}$ and PTS correlations.

The PTS length scale is extracted by calculating the range over which spatial correlations imposed by an equilibrium amorphous spherical boundary decay. We first establish that it is possible to ergodically sample cavities at some $R_{cav}$ using the ``Particle Size Annealing'' (PSA) method detailed in Ref.~\onlinecite{Hocky-PRL2012}.
In brief, we monitor the overlap $q$, a measure of the similarity between the initial configuration in the cavity and that at a later time $t$.
The overlap is defined as $q(R_{cav},t)=(\rho l^3 \tilde N)^{-1} \sum_{i=1}^{\tilde{N}} \langle n_i(t)n_i(0) \rangle$ where the center of the cavity has been tiled into $\tilde{N}=125$ cubes of side length $l=0.36$, small enough such that the cell occupancy $n_i(t)$ is always zero or one.
We use both regular Monte Carlo (MC) sampling and a sampling where the particle diameters are shrunk to 60\% of their original size and grown back in, and check that the $q$ values agree at long times. 
In the limit of large cavity and long time, $q(t)$ will tend to the bulk value $q_b=\rho l^3$ and so this value is conventionally subtracted from $q$.

\begin{figure}[t]
\centering
\includegraphics[width=3.4in]{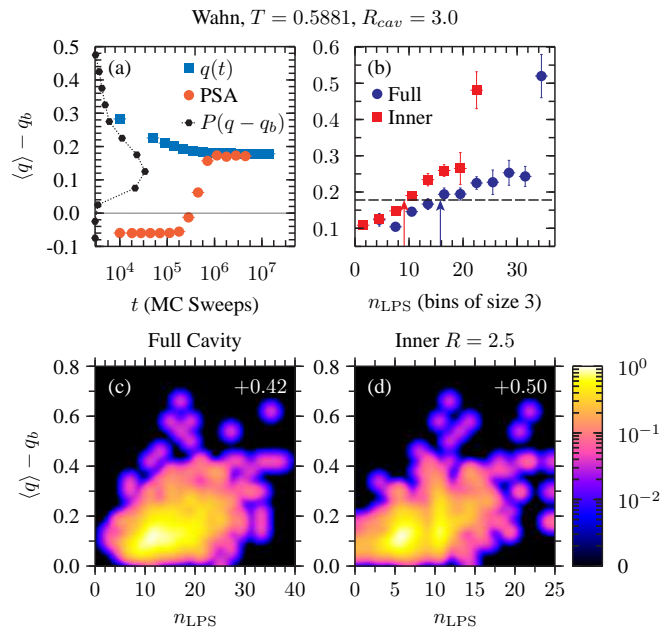}
\caption{(a) Cavity overlaps and PSA overlaps (with the value for a bulk system $q_b$ subtracted) indicating ergodic sampling at this cavity size and temperature. The dashed line shows the overlap probability distribution $P(q-q_b)$. (b) On average, cavities containing a large number of LPS centers have high overlap.
``Inner'' points count only LPS which are within $R=2.5$ of the center of the cavity. The dashed line shows the average overlap and arrows show the average number of LPS. (c) and (d) show the full data distribution and Spearman rank correlation coefficients~\cite{Sammut-Encyclopedia2011}. The data in (b) is obtained from (c) and (d) by averaging over vertical slabs of width 3.}
\label{fig:overlap_corr}
\end{figure}

We carry out these tests for the Wahn model. The strong icosahedral ordering in this model makes it an ideal system to probe the connection between local order and the spatial distribution of the overlap. In Fig.~\ref{fig:overlap_corr}(a) we show that for $R_{cav}=3.0$, $q$ is sampled ergodically.
We then take 30 of the Wahn configurations used earlier and perform two standard MC simulations for cavities centered at $27$ positions in each.
The longtime overlap value is extracted from each cavity, and the number of icosahedral centers within the cavity as well as the one within the inner $R=2.5$ of the cavity, are calculated.
We find that high overlap cavities generally have high $n_{\rm LPS}$ and {\em vice versa}.
This implies that (for the Wahn system) the cavity simulations are mostly probing the same type of local ordering measured by the Voronoi construction, although it does not necessarily mean that the correlation {\em length} measured by doing cavity simulations at a series of radii is the same as would be measured by the extent of LPS domains.

In conclusion, we have demonstrated that the correlation between local structural metrics (e.g. $E_i$ and $n_{\rm LPS}$) and dynamics in supercooled liquids is highly system dependent. In models such as the Wahn mixture, accurate predictions of heterogeneous dynamics may be made on the basis of a single structural marker while essentially no correlation exists in mean-field like systems such as the GCM. However, a strong link between {\em some} aspect of static structure and dynamics does exist, as signified by the sizable $R_4$-ratio observed in all the systems we have studied. Despite being order agnostic, PTS correlations appear to show a connection with specific types of local order such as Voronoi signatures in systems whose dynamics may be predicted by the location of such structural motifs. Furthermore, previous work has detected an apparent connection between the growth of relaxation times and order agnostic length scales in systems where the connection between relaxation times and specific structural metrics is not very strong~\cite{Hocky-PRL2012,Biroli-PRL2013,Karmakar-ARCMP2014}. These facts suggest that PTS correlations may provide a more general description of the key static fluctuations that determine dynamical behavior in supercooled liquids. 

\begin{acknowledgments}
We thank Mark Ediger and Asaph Widmer-Cooper for stimulating discussions. Simulations were executed in part on the Midway cluster at the University of Chicago's Research Computing Center, and on the seeder cluster of the University of Chicago Computing Cooperative (UC3), supported in part by the Open Science Grid, NSF Grant No. PHY-1148698. LAMMPS~\cite{Plimpton-JCP1995} simulations were organized and executed using the Swift parallel scripting language, development supported by NSF Grant No.~OCI-1148443~\cite{Wilde-Parallel2011}. G.M.H. and D.R.R. were supported by the NSF through Grant No. DGE-07-07425 and Grant No. CHE-1213247, respectively.
\end{acknowledgments}


%
\begin{appendix}

\onecolumngrid
\newpage
\section{Supplemental Information}
\setcounter{page}{1}
\label{sec:sup}

\begin{table}[h]
\begin{ruledtabular}
\begin{tabular}{c c c c  c c c c c }
System & $N$ &$\rho$ & $T$ & $T_c$ & $k$ & $\delta_{4}(\tau_\alpha)$ & $\chi_{4}(\tau_\alpha)$ & $R_{4}(\tau_\alpha)$ \\
\hline
Wahn &1000& 1.297 & 0.588 &0.56~\supcite{Kim_Saito_2013} & 7.7 & 5.60 & 8.99 & 0.623\\
KA & 1000& 1.2& 0.45 &0.435~\supcite{Kob-PRL1994} & 7.25 &  13.3 & 21.4 & 0.620 \\
Harm &1000& 0.675&5.5 &5.2~\supcite{Kob-NatPhys2011} & 6.28 & 3.61 & 9.20 & 0.392\\
GCM &3456& 2.0&3.2 &2.7~\supcite{Ikeda-JCP2011} & 8.40 & 8.08 & 18.9 & 0.428
\end{tabular}
\end{ruledtabular}
\caption{Predictability ratio results for the four systems studied. See Ref.~\supcite{Berthier-PRE2007} and SI text for further discussion.}
\label{tab:predict-ratio}
\end{table} 

\subsection{Models, units and relevant temperatures}

The Kob-Andersen (KA)~\supcite{Kob-PRL1994} and Wahnstr\"om
(Wahn)~\supcite{Wahnstrom-PRA1991} models are binary Lennard-Jones
mixtures where particles interact through the potential
\begin{equation} 
u_{\alpha\beta}(r) = 4\epsilon_{\alpha\beta}\left[
  {\left( \frac{\sigma_{\alpha\beta}}{r} \right)}^{12} - {\left(
    \frac{\sigma_{\alpha\beta}}{r} \right)}^6 \right],
\end{equation}
where $\alpha, \beta = A,B$ are species indices. For the KA model the
interaction parameters are
$\sigma_{AB}=0.8\sigma_{AA}$, $\sigma_{BB}=0.88\sigma_{AA}$,
$\epsilon_{AB}=1.5\epsilon_{AA}$, and $\epsilon_{BB}=0.5\epsilon_{AA}$, while for the Wahn
model $\sigma_{AB}=0.916\sigma_{AA}$, $\sigma_{BB}=0.833\sigma_{AA}$, and
$\epsilon_{BB}=\epsilon_{AB}=\epsilon_{AA}$. The
chemical composition is $x_1=1-x_2=0.8$ for the KA model 
and $x_1=x_2=0.5$ for the Wahn model.
The mass ratio $m_1/m_2$ is 1 and 2 in KA and Wahn, respectively.
The potentials are cut and shifted at $2.5\sigma_{\alpha\beta}$. 
In the main text, $\sigma_{AA}$, $\epsilon_{AA}$, and
$\sqrt{m_A\sigma_{AA}^2/\epsilon_{AA}}$ are used as units of distance, energy,
and time, respectively. 

The Harm model is an equimolar mixture of elastic spheres~\supcite{OHern_Langer_Liu_Nagel_2002} with equal masses $m$.
The interaction potential is given by
\begin{equation}
u_{\alpha\beta}(r) = \frac{\epsilon}{2} \left(1-\frac{r}{\sigma_{\alpha\beta}}\right)^2,
\end{equation}
if $r_{\alpha\beta} < \sigma_{\alpha\beta}$ and zero otherwise. 
The interaction parameters are $\sigma_{AA}=1$, $\sigma_{AB}=1.2$ and $\sigma_{BB}=1.4$. 
As in the KA and Wahn models, $\sigma_{AA}$ and $\sqrt{m\sigma_{AA}^2/\epsilon}$ are used as units of length and time, respectively. The unit of energy is $10^{-4}\epsilon$.

The Gaussian core model~\cite{Stillinger_1976-sup,Lang_Likos_Watzlawek_Lowen_2000-sup} (GCM) is a one-component fluid of particles interacting via a Gaussian potential
$$
u(r) = \epsilon e^{-(r/\sigma)^2}.
$$
The potential is cut and shifted at $r=5\sigma$.
The units of length, energy and time are given by $\sigma$, $10^{-6}\epsilon$ and $\sqrt{m\sigma^2/\epsilon}$ respectively.

The simulations have been carried out using the LAMMPS package~\supcite{Plimpton-JCP1995}. Integration time steps $\delta t$ used were 0.004 for the Wahn and KA system, 0.1 for Harm, and 0.4 for the GCM. Temperature was maintained by a Nos\'{e}-Hoover thermostat with a time constant of 100 $\delta t$ in all cases.

The estimated Mode-Coupling critical temperatures of all the models and the wave-vectors $k$ used to compute the structural relaxation times $\tau_\alpha$ (see section on $R_4$ values) are reported in Tab.~\ref{tab:predict-ratio}. The estimated  critical temperatures were obtained, in the references cited in in Tab.~\ref{tab:predict-ratio}, from power law fits to the relaxation times data.

\subsection{Identification of locally preferred structures}
Particles participating in LPS are identified through a Voronoi analysis~\supcite{Coslovich-PRE2011}.
To characterize the local structure around a given particle, we determine the number $n_k$ of faces having $k$ edges of the Voronoi polyhedron formed by the nearest neighbors.
Each particle may therefore be labeled by a Voronoi signature (VS) $(n_3,n_4,n_5,..)$, where $n_3$ is the number of triangles, $n_4$ the number of quadrilaterals, etc. of the corresponding Voronoi polyhedron.
As in Ref.~\onlinesupcite{Coslovich-PRE2011}, we focus on the polyhedra found around small particles (for binary mixtures), whose temperature variation were shown to correlate better with slow dynamics.
In the Wahn system, the $(0,0,12)$ polyhedron (icosahedron) appears around over 25\% of type $B$ particles at low temperatures, while the second most common VS, $(0,2,8,1)$, appears around approximately 9\% of $B$ particles.
In the KA system, the $(0,2,8)$ arrangement (bicapped square antiprism) appears around approximately 10\% of small particles and the $(1,2,5,3)$ appear around 8\%.
It is important to remember that, because the KA system is an 80:20 mixture, the $(0,2,8)$ motif is found in only about 2\% abundance while more than 10\% of Wahn particles are found in $(0,0,12)$ configurations. 
We therefore considered the union of $(0,2,8)$ and $(1,2,5,3)$ polyhedra as the LPS of KA model.
The analysis carried out in the main text remains qualitatively unchanged if only $(0,2,8)$ are considered.
In the Harm system, the most frequent signature at low temperature is $(0,2,8,2)$, which may be regarded as a distorted icosahedral structure. 
The abundance of this LPS is about 5\% at the lowest studied temperatures, while the second most frequent signature is $(0,2,8,1)$ has 4\%.
In the GCM, we found a variety of low symmetry polyhedra, some of which are found in FCC and BCC crystals at finite temperatures~\supcite{tanemura_geometrical_1977}. At the studied temperature, $(0,3,6,4)$, $(0,2,8,4)$, and $(0,4,4,6)$ (ideal FCC structure) are the most abundant ones. These crystal-like structures tend to grow slightly by decreasing temperature and their union is considered as the LPS of the model.

\subsection{Mean-field behavior and pair potential}
\begin{figure}[ht]
\includegraphics{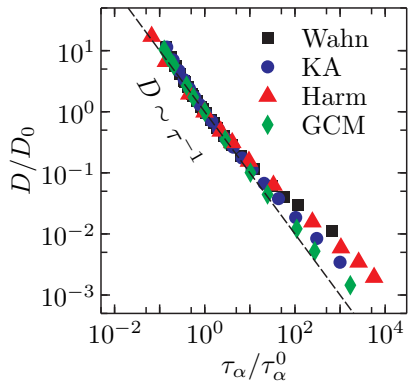}
\caption{Diffusion constants vs. $\tau_\alpha$ (see SI text) for each of the four models. Open symbols represent the temperatures studied in the main body of this work. The diffusion constants $D$ were extracted from the limiting relationship $\lim_{t\rightarrow \infty} \delta \vec{r}^2_i(t) = 6 D t$. The values of the diffusion constants and $\tau_\alpha$ are scaled by their value at the approximate onset temperature of glassy behavior, with $T_0$ values used here of 1.0, 1.0, 12 and 5 for the Wahn, KA, Harm and GCM respectively. The dashed line shows the approximate high temperature behavior $D\propto\tau^{-1}$.}
\label{fig:se}
\end{figure}

As mentioned in the body of the text, the Harm and GCM systems seem to embody manifestations of the predictions of mean-field theories of glasses, however they do so in distinct ways. The Harm system shows a well defined crossover between seemingly activation-less and activated regimes as predicted by the mean-field perspective. However, SE violation is still pronounced in the Harm system (see Fig.~\ref{fig:se}). In this sense, critical fluctuations expected in finite dimensions \supcite{Biroli-JPCM2007} are still sizable. In the GCM system, the entire range of temperatures that can be simulated displays a near-complete, quantifiable correspondence with high dimensional mean-field behavior, including the absence of Stokes-Einstein violation (despite the GCM system being highly fragile). Both potentials share an ultra-soft core, which prevents the physics from being dominated by harsh short-ranged repulsions. 
In addition, the GCM potential is (quasi) long-ranged, mimicking the effect of increased dimensionality \cite{Eaves-PNAS2009-sup,Charbonneau-PRE2010-sup,Ikeda-PRL2011-sup,Charbonneau-JCP2013-sup}. 
Thus ultra-softness may be a necessary condition for triggering aspects of mean-field behavior. However, to further suppress critical-like fluctuations in low dimensions, a (quasi) long-ranged potential, as found in the GCM, is also necessary. 

\subsection{$R_{4}$ values}
\label{sec:r4}
In order to quantify the total connection between structure and dynamics we perform the analysis of Berthier and Jack (Ref.~\supcite{Berthier-PRE2007}). They definite the quantity $R_{4}(t)$ which represents the fraction of the total run-to-run variance in a dynamical quantity from analyzed in the isoconfigurational ensemble which is encoded by the initial configurations. Hence a value of zero means all fluctuations come from the randomly selected initial velocities while a value of unity means that all fluctuations in the dynamical quantity are determined by the initial structure. $R_4$ is defined as,
\begin{equation}
R_{4}(t)=\frac{ \delta_{4}(t) }{\chi_{4}(t)}, 
\end{equation}
where 
\begin{eqnarray}
\delta_{4}(t) = N \{\mathbb{E} [ \isoavg{F(t)}^2]-\mathbb{E}[F(t)]^2\}\\
\chi_{4}(t) = N \{\mathbb{E} [ \isoavg{F^2(t)}]-\mathbb{E}[F(t)]^2\}.
\end{eqnarray}
Here, ``F'' is a collective dynamical quantity, a function averaged over all particles. $\mathbb{E}$ represents an equilibrium ensemble average while $\isoavg{\cdots}$ represents the isoconfigurational average over realizations of momenta. Here we choose $F(t)$ to be the self-intermediate scattering function $F_s(k,t)=\frac{1}{N} \sum_{i=1}^{N} e^{\vec{k}\cdot\delta \vec{r}_i(t)}$. The relaxation time $\tau_\alpha$ is defined by $F_s(k,t\equiv\tau_\alpha)=1/e$. For the four systems studied in this work, we report the results of the $R_{4}$ analysis at $t=\tau_\alpha$ in Tab.~\ref{tab:predict-ratio}. For all four systems, we find values that should be considered large based on prior analyses using this quantity, though as would be expected from our other results and discussion, it is smaller in the Harm and GCM systems than in the two LJ systems. 

\subsection{Mobility correlation and predictability}
\begin{figure}[ht]
\includegraphics[width=3.2in]{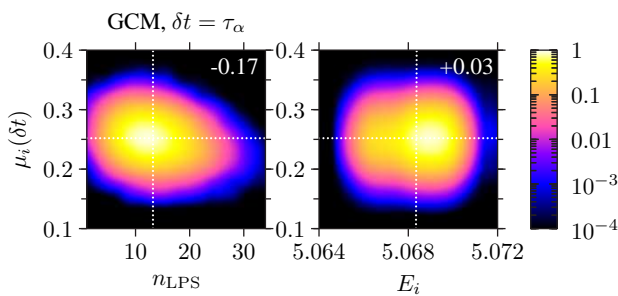}
\caption{Mobility $\mu_i$ and local energy $E_i$ vs. $n_{\rm LPS}$ histogram for the GCM as in Fig.~\ref{fig:lpshist}.}
\label{fig:gcmhist}
\end{figure}

\begin{figure}[ht]
\includegraphics[width=2.8in]{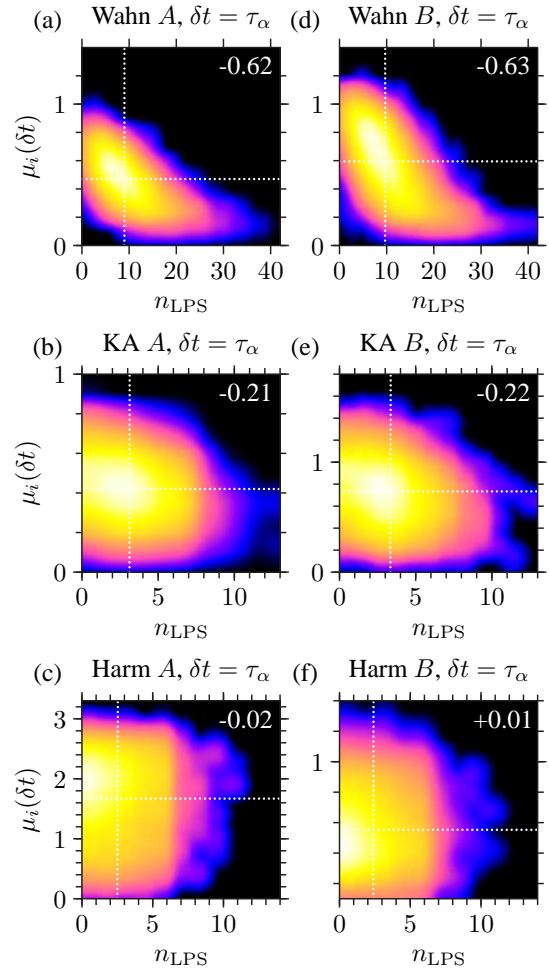}
\caption{Mobility $\mu_i$ vs. $n_{\rm LPS}$ histogram for the three models in Fig.~\ref{fig:lpshist} comparing both species of particles. The larger particle always has a smaller range of dynamical activity (note the differing ranges of vertical axes) and this intrinsically reduces the correlation.}
\label{fig:lpshist-ab}
\end{figure}

\begin{figure*}[ht]
\includegraphics[scale=1.0]{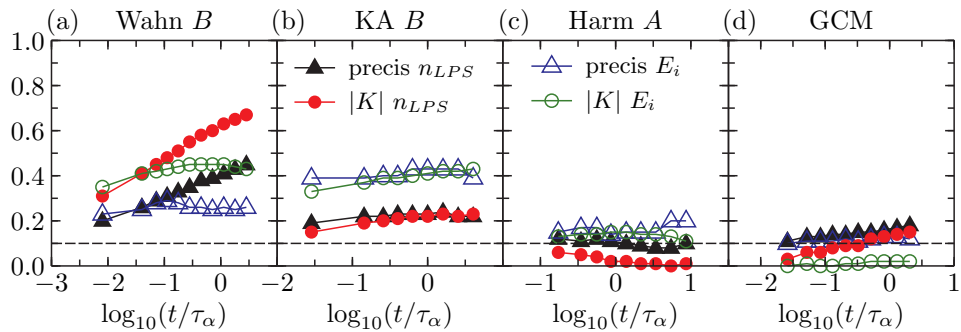}
\caption{Same as Fig.~\ref{fig:predict} except the precision in that figure, the precision of prediction the slowest 10\% of particles, has been replaced by the precision of predicting the fastest 10\% of particles based on a lack of local stability (high $E_i$, small $n_{\rm LPS}$). We see that the trends are all the same as for the ``slow precision'', however, the ``fast precision'' is much worse for both $n_{LPS}$ and $E_i$ in the Wahn model, reflecting the wider density at the left and than on the right end in Fig.~\ref{fig:lpshist}. In contrast, the precision for the KA system does not change much reflecting the more ovular shape of the density in Fig.~\ref{fig:lpshist}.}
\label{fig:predict-fast}
\end{figure*}

For completeness, in Fig.~\ref{fig:gcmhist} we present histograms for the GCM data analogous to that of Fig.~\ref{fig:lpshist}. In Fig.~\ref{fig:lpshist-ab} we show the histograms for both the $A$ and $B$ types of particles for the three binary systems presented in Fig.~\ref{fig:lpshist}.
In Fig.~\ref{fig:predict-fast} we show the results for predictability of the fastest 10\% of particles based on the lowest 2\% of $n_{\rm LPS}$ or the highest 2\% of $E_i$ for contrast with Fig.~\ref{fig:predict}. For comparison, the same data for $K$ as in Fig.~\ref{fig:predict} are also shown.

\begin{figure*}[ht]
\includegraphics[scale=1.0]{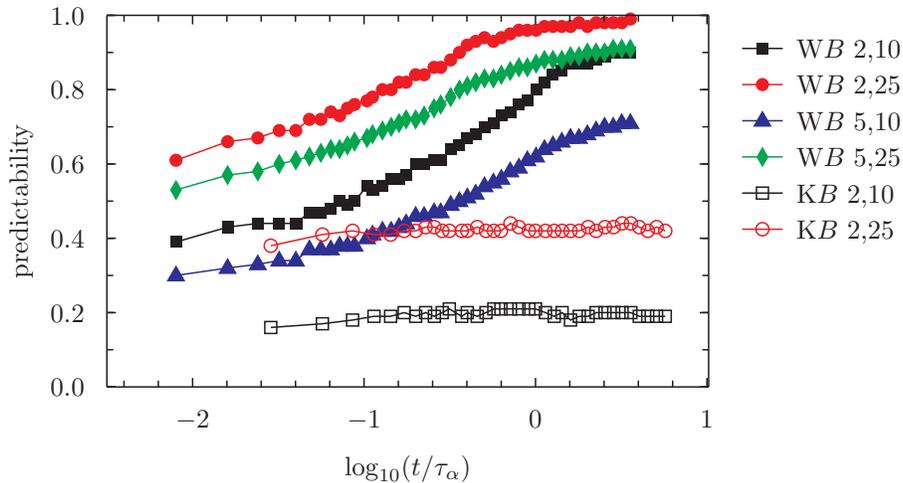}
\caption{Predictability of mobility from $n_{LPS}$ as a function of time, as in Fig.~\ref{fig:predict}. Data is shown for the Wahn (closed symbols) and KA (open symbols) $B$ particles, with the definition of predictability varied. In the legend, the first symbol represents the cutoff percentile for $n_{LPS}$ and the second number for mobility. Hence the data in the main text is the same as that labeled ``2,10''. For the case of the KA system, due to the smaller number of particles involved in LPS, the particles in the top five percent in $n_{LPS}$ are precisely those in the second percentile, hence the other set of data lies on top of those shown.}
\label{fig:predict-compare-def}
\end{figure*}

We show in Fig.~\ref{fig:predict-compare-def} how the predictability data is modified by a change in how predictability is defined. For simplicity, we illustrate the point with data for the Wahn $B$ and KA $B$ particles and show how, while the values change, the trends discussed in the main text are robust.

Lastly, we note that there can be large sample-to-sample variations in $n_{LPS}$ observed around particles at the large $n_{LPS}$ end. We note that the percentile thresholds used in this calculation are done on the aggregate set of data, which means that the number of particles exceeding this threshold can vary substantially from sample to sample. As an example, we find in the Wahn system that the number of $B$ particles with $n_{LPS}$ above the 2\% threshold ($n_{LPS}\geq 23$) varies for our forty samples from zero to eighty out of 500 $B$ particles. Nevertheless, we feel this strengthens our point that for the Wahnstrom system, very large numbers of locally preferred structures in an area is highly predictive of slow dynamics, although in general slow dynamics can also arise via other {\em structural} mechanisms, as suggested by our $R_4$ analysis above.

%
\end{appendix}


\begin{thebibliography}{54}%
\makeatletter
\providecommand \@ifxundefined [1]{%
 \@ifx{#1\undefined}
}%
\providecommand \@ifnum [1]{%
 \ifnum #1\expandafter \@firstoftwo
 \else \expandafter \@secondoftwo
 \fi
}%
\providecommand \@ifx [1]{%
 \ifx #1\expandafter \@firstoftwo
 \else \expandafter \@secondoftwo
 \fi
}%
\providecommand \natexlab [1]{#1}%
\providecommand \enquote  [1]{``#1''}%
\providecommand \bibnamefont  [1]{#1}%
\providecommand \bibfnamefont [1]{#1}%
\providecommand \citenamefont [1]{#1}%
\providecommand \href@noop [0]{\@secondoftwo}%
\providecommand \href [0]{\begingroup \@sanitize@url \@href}%
\providecommand \@href[1]{\@@startlink{#1}\@@href}%
\providecommand \@@href[1]{\endgroup#1\@@endlink}%
\providecommand \@sanitize@url [0]{\catcode `\\12\catcode `\$12\catcode
  `\&12\catcode `\#12\catcode `\^12\catcode `\_12\catcode `\%12\relax}%
\providecommand \@@startlink[1]{}%
\providecommand \@@endlink[0]{}%
\providecommand \url  [0]{\begingroup\@sanitize@url \@url }%
\providecommand \@url [1]{\endgroup\@href {#1}{\urlprefix }}%
\providecommand \urlprefix  [0]{URL }%
\providecommand \Eprint [0]{\href }%
\providecommand \doibase [0]{http://dx.doi.org/}%
\providecommand \selectlanguage [0]{\@gobble}%
\providecommand \bibinfo  [0]{\@secondoftwo}%
\providecommand \bibfield  [0]{\@secondoftwo}%
\providecommand \translation [1]{[#1]}%
\providecommand \BibitemOpen [0]{}%
\providecommand \bibitemStop [0]{}%
\providecommand \bibitemNoStop [0]{.\EOS\space}%
\providecommand \EOS [0]{\spacefactor3000\relax}%
\providecommand \BibitemShut  [1]{\csname bibitem#1\endcsname}%
\let\auto@bib@innerbib\@empty
\bibitem [{\citenamefont {Ediger}(2000)}]{Ediger-ARPC2000}%
  \BibitemOpen
  \bibfield  {author} {\bibinfo {author} {\bibfnamefont {M.~D.}\ \bibnamefont
  {Ediger}},\ }\href@noop {} {\bibfield  {journal} {\bibinfo  {journal} {Annu.
  Rev. Phys. Chem.}\ }\textbf {\bibinfo {volume} {51}},\ \bibinfo {pages} {99}
  (\bibinfo {year} {2000})}\BibitemShut {NoStop}%
\bibitem [{\citenamefont {Berthier}\ \emph {et~al.}(2011)\citenamefont
  {Berthier}, \citenamefont {Biroli}, \citenamefont {Bouchaud}, \citenamefont
  {Cipelletti},\ and\ \citenamefont {van Saarloos}}]{Berthier-Book2011}%
  \BibitemOpen
  \bibfield  {author} {\bibinfo {author} {\bibfnamefont {L.}~\bibnamefont
  {Berthier}}, \bibinfo {author} {\bibfnamefont {G.}~\bibnamefont {Biroli}},
  \bibinfo {author} {\bibfnamefont {J.-P.}\ \bibnamefont {Bouchaud}}, \bibinfo
  {author} {\bibfnamefont {L.}~\bibnamefont {Cipelletti}}, \ and\ \bibinfo
  {author} {\bibfnamefont {W.}~\bibnamefont {van Saarloos}},\ }\href@noop {}
  {\emph {\bibinfo {title} {Dynamical heterogeneities in glasses, colloids, and
  granular media}}}\ (\bibinfo  {publisher} {Oxford University Press},\
  \bibinfo {year} {2011})\BibitemShut {NoStop}%
\bibitem [{\citenamefont {Berthier}\ and\ \citenamefont
  {Biroli}(2011)}]{Berthier-RMP2011}%
  \BibitemOpen
  \bibfield  {author} {\bibinfo {author} {\bibfnamefont {L.}~\bibnamefont
  {Berthier}}\ and\ \bibinfo {author} {\bibfnamefont {G.}~\bibnamefont
  {Biroli}},\ }\href@noop {} {\bibfield  {journal} {\bibinfo  {journal} {Rev.
  Mod. Phys.}\ }\textbf {\bibinfo {volume} {83}},\ \bibinfo {pages} {587}
  (\bibinfo {year} {2011})}\BibitemShut {NoStop}%
\bibitem [{\citenamefont {Widmer-Cooper}\ \emph {et~al.}(2004)\citenamefont
  {Widmer-Cooper}, \citenamefont {Harrowell},\ and\ \citenamefont
  {Fynewever}}]{WidmerCooper-PRL2004}%
  \BibitemOpen
  \bibfield  {author} {\bibinfo {author} {\bibfnamefont {A.}~\bibnamefont
  {Widmer-Cooper}}, \bibinfo {author} {\bibfnamefont {P.}~\bibnamefont
  {Harrowell}}, \ and\ \bibinfo {author} {\bibfnamefont {H.}~\bibnamefont
  {Fynewever}},\ }\href@noop {} {\bibfield  {journal} {\bibinfo  {journal}
  {Phys. Rev. Lett.}\ }\textbf {\bibinfo {volume} {93}},\ \bibinfo {pages}
  {135701} (\bibinfo {year} {2004})}\BibitemShut {NoStop}%
\bibitem [{\citenamefont {Widmer-Cooper}\ and\ \citenamefont
  {Harrowell}(2005)}]{WidmerCooper-JPCM2005}%
  \BibitemOpen
  \bibfield  {author} {\bibinfo {author} {\bibfnamefont {A.}~\bibnamefont
  {Widmer-Cooper}}\ and\ \bibinfo {author} {\bibfnamefont {P.}~\bibnamefont
  {Harrowell}},\ }\href@noop {} {\bibfield  {journal} {\bibinfo  {journal} {J.
  Phys: Cond. Mat.}\ }\textbf {\bibinfo {volume} {17}},\ \bibinfo {pages}
  {S4025} (\bibinfo {year} {2005})}\BibitemShut {NoStop}%
\bibitem [{\citenamefont {Widmer-Cooper}\ and\ \citenamefont
  {Harrowell}(2006)}]{WidmerCooper-JNCS2006}%
  \BibitemOpen
  \bibfield  {author} {\bibinfo {author} {\bibfnamefont {A.}~\bibnamefont
  {Widmer-Cooper}}\ and\ \bibinfo {author} {\bibfnamefont {P.}~\bibnamefont
  {Harrowell}},\ }\href@noop {} {\bibfield  {journal} {\bibinfo  {journal} {J.
  Non-Cryst. Solids}\ }\textbf {\bibinfo {volume} {352}},\ \bibinfo {pages}
  {5098} (\bibinfo {year} {2006})}\BibitemShut {NoStop}%
\bibitem [{\citenamefont {Widmer-Cooper}\ \emph {et~al.}(2008)\citenamefont
  {Widmer-Cooper}, \citenamefont {Perry}, \citenamefont {Harrowell},\ and\
  \citenamefont {Reichman}}]{WidmerCooper-NatPhys2008}%
  \BibitemOpen
  \bibfield  {author} {\bibinfo {author} {\bibfnamefont {A.}~\bibnamefont
  {Widmer-Cooper}}, \bibinfo {author} {\bibfnamefont {H.}~\bibnamefont
  {Perry}}, \bibinfo {author} {\bibfnamefont {P.}~\bibnamefont {Harrowell}}, \
  and\ \bibinfo {author} {\bibfnamefont {D.~R.}\ \bibnamefont {Reichman}},\
  }\href@noop {} {\bibfield  {journal} {\bibinfo  {journal} {Nat. Phys.}\
  }\textbf {\bibinfo {volume} {4}},\ \bibinfo {pages} {711} (\bibinfo {year}
  {2008})}\BibitemShut {NoStop}%
\bibitem [{\citenamefont {Widmer-Cooper}\ \emph {et~al.}(2009)\citenamefont
  {Widmer-Cooper}, \citenamefont {Perry}, \citenamefont {Harrowell},\ and\
  \citenamefont {Reichman}}]{WidmerCooper-JCP2009}%
  \BibitemOpen
  \bibfield  {author} {\bibinfo {author} {\bibfnamefont {A.}~\bibnamefont
  {Widmer-Cooper}}, \bibinfo {author} {\bibfnamefont {H.}~\bibnamefont
  {Perry}}, \bibinfo {author} {\bibfnamefont {P.}~\bibnamefont {Harrowell}}, \
  and\ \bibinfo {author} {\bibfnamefont {D.~R.}\ \bibnamefont {Reichman}},\
  }\href@noop {} {\bibfield  {journal} {\bibinfo  {journal} {J. Chem. Phys.}\
  }\textbf {\bibinfo {volume} {131}},\ \bibinfo {pages} {194508} (\bibinfo
  {year} {2009})}\BibitemShut {NoStop}%
\bibitem [{\citenamefont {Candelier}\ \emph {et~al.}(2010)\citenamefont
  {Candelier}, \citenamefont {Widmer-Cooper}, \citenamefont {Kummerfeld},
  \citenamefont {Dauchot}, \citenamefont {Biroli}, \citenamefont {Harrowell},\
  and\ \citenamefont {Reichman}}]{Candelier-PRL2010}%
  \BibitemOpen
  \bibfield  {author} {\bibinfo {author} {\bibfnamefont {R.}~\bibnamefont
  {Candelier}}, \bibinfo {author} {\bibfnamefont {A.}~\bibnamefont
  {Widmer-Cooper}}, \bibinfo {author} {\bibfnamefont {J.~K.}\ \bibnamefont
  {Kummerfeld}}, \bibinfo {author} {\bibfnamefont {O.}~\bibnamefont {Dauchot}},
  \bibinfo {author} {\bibfnamefont {G.}~\bibnamefont {Biroli}}, \bibinfo
  {author} {\bibfnamefont {P.}~\bibnamefont {Harrowell}}, \ and\ \bibinfo
  {author} {\bibfnamefont {D.~R.}\ \bibnamefont {Reichman}},\ }\href@noop {}
  {\bibfield  {journal} {\bibinfo  {journal} {Phys. Rev. Lett.}\ }\textbf
  {\bibinfo {volume} {105}},\ \bibinfo {pages} {135702} (\bibinfo {year}
  {2010})}\BibitemShut {NoStop}%
\bibitem [{\citenamefont {Coslovich}\ and\ \citenamefont
  {Pastore}(2006)}]{Coslovich_Pastore_2006}%
  \BibitemOpen
  \bibfield  {author} {\bibinfo {author} {\bibfnamefont {D.}~\bibnamefont
  {Coslovich}}\ and\ \bibinfo {author} {\bibfnamefont {G.}~\bibnamefont
  {Pastore}},\ }\href@noop {} {\bibfield  {journal} {\bibinfo  {journal}
  {Europhys. Lett.}\ }\textbf {\bibinfo {volume} {75}},\ \bibinfo {pages} {784}
  (\bibinfo {year} {2006})}\BibitemShut {NoStop}%
\bibitem [{\citenamefont {Frank}(1952)}]{Frank-PRS1952}%
  \BibitemOpen
  \bibfield  {author} {\bibinfo {author} {\bibfnamefont {F.~C.}\ \bibnamefont
  {Frank}},\ }\href@noop {} {\bibfield  {journal} {\bibinfo  {journal} {Proc.
  R. Soc. Lond., A}\ }\textbf {\bibinfo {volume} {215}},\ \bibinfo {pages} {43}
  (\bibinfo {year} {1952})}\BibitemShut {NoStop}%
\bibitem [{\citenamefont {Steinhardt}\ \emph {et~al.}(1981)\citenamefont
  {Steinhardt}, \citenamefont {Nelson},\ and\ \citenamefont
  {Ronchetti}}]{Steinhardt-PRL1981}%
  \BibitemOpen
  \bibfield  {author} {\bibinfo {author} {\bibfnamefont {P.~J.}\ \bibnamefont
  {Steinhardt}}, \bibinfo {author} {\bibfnamefont {D.~R.}\ \bibnamefont
  {Nelson}}, \ and\ \bibinfo {author} {\bibfnamefont {M.}~\bibnamefont
  {Ronchetti}},\ }\href@noop {} {\bibfield  {journal} {\bibinfo  {journal}
  {Phys. Rev. Lett.}\ }\textbf {\bibinfo {volume} {47}},\ \bibinfo {pages}
  {1297} (\bibinfo {year} {1981})}\BibitemShut {NoStop}%
\bibitem [{\citenamefont {Nelson}(2002)}]{Nelson-2002Defects}%
  \BibitemOpen
  \bibfield  {author} {\bibinfo {author} {\bibfnamefont {D.~R.}\ \bibnamefont
  {Nelson}},\ }\href@noop {} {\emph {\bibinfo {title} {Defects and geometry in
  condensed matter physics}}}\ (\bibinfo  {publisher} {Cambridge University
  Press},\ \bibinfo {year} {2002})\BibitemShut {NoStop}%
\bibitem [{\citenamefont {Kivelson}\ \emph {et~al.}(1995)\citenamefont
  {Kivelson}, \citenamefont {Kivelson}, \citenamefont {Zhao}, \citenamefont
  {Nussinov},\ and\ \citenamefont {Tarjus}}]{Kivelson-PhysicaA1995}%
  \BibitemOpen
  \bibfield  {author} {\bibinfo {author} {\bibfnamefont {D.}~\bibnamefont
  {Kivelson}}, \bibinfo {author} {\bibfnamefont {S.~A.}\ \bibnamefont
  {Kivelson}}, \bibinfo {author} {\bibfnamefont {X.}~\bibnamefont {Zhao}},
  \bibinfo {author} {\bibfnamefont {Z.}~\bibnamefont {Nussinov}}, \ and\
  \bibinfo {author} {\bibfnamefont {G.}~\bibnamefont {Tarjus}},\ }\href@noop {}
  {\bibfield  {journal} {\bibinfo  {journal} {Physica A}\ }\textbf {\bibinfo
  {volume} {219}},\ \bibinfo {pages} {27} (\bibinfo {year} {1995})}\BibitemShut
  {NoStop}%
\bibitem [{\citenamefont {Coslovich}\ and\ \citenamefont
  {Pastore}(2007)}]{Coslovich-JCP2007}%
  \BibitemOpen
  \bibfield  {author} {\bibinfo {author} {\bibfnamefont {D.}~\bibnamefont
  {Coslovich}}\ and\ \bibinfo {author} {\bibfnamefont {G.}~\bibnamefont
  {Pastore}},\ }\href@noop {} {\bibfield  {journal} {\bibinfo  {journal} {J.
  Chem. Phys.}\ }\textbf {\bibinfo {volume} {127}},\ \bibinfo {pages} {124504}
  (\bibinfo {year} {2007})}\BibitemShut {NoStop}%
\bibitem [{\citenamefont {Ding}\ \emph {et~al.}(2012)\citenamefont {Ding},
  \citenamefont {Cheng}, \citenamefont {Sheng},\ and\ \citenamefont
  {Ma}}]{Ding_Cheng_Sheng_Ma_2012}%
  \BibitemOpen
  \bibfield  {author} {\bibinfo {author} {\bibfnamefont {J.}~\bibnamefont
  {Ding}}, \bibinfo {author} {\bibfnamefont {Y.-Q.}\ \bibnamefont {Cheng}},
  \bibinfo {author} {\bibfnamefont {H.}~\bibnamefont {Sheng}}, \ and\ \bibinfo
  {author} {\bibfnamefont {E.}~\bibnamefont {Ma}},\ }\href@noop {} {\bibfield
  {journal} {\bibinfo  {journal} {Phys. Rev. B}\ }\textbf {\bibinfo {volume}
  {85}},\ \bibinfo {pages} {060201} (\bibinfo {year} {2012})}\BibitemShut
  {NoStop}%
\bibitem [{\citenamefont {Tanaka}\ \emph {et~al.}(2010)\citenamefont {Tanaka},
  \citenamefont {Kawasaki}, \citenamefont {Shintani},\ and\ \citenamefont
  {Watanabe}}]{Tanaka_Kawasaki_Shintani_Watanabe_2010}%
  \BibitemOpen
  \bibfield  {author} {\bibinfo {author} {\bibfnamefont {H.}~\bibnamefont
  {Tanaka}}, \bibinfo {author} {\bibfnamefont {T.}~\bibnamefont {Kawasaki}},
  \bibinfo {author} {\bibfnamefont {H.}~\bibnamefont {Shintani}}, \ and\
  \bibinfo {author} {\bibfnamefont {K.}~\bibnamefont {Watanabe}},\ }\href
  {\doibase 10.1038/nmat2634} {\bibfield  {journal} {\bibinfo  {journal}
  {Nature Mater.}\ }\textbf {\bibinfo {volume} {9}},\ \bibinfo {pages} {324}
  (\bibinfo {year} {2010})}\BibitemShut {NoStop}%
\bibitem [{\citenamefont {Malins}\ \emph {et~al.}(2013)\citenamefont {Malins},
  \citenamefont {Eggers}, \citenamefont {Royall}, \citenamefont {Williams},\
  and\ \citenamefont {Tanaka}}]{Malins-JCP2013}%
  \BibitemOpen
  \bibfield  {author} {\bibinfo {author} {\bibfnamefont {A.}~\bibnamefont
  {Malins}}, \bibinfo {author} {\bibfnamefont {J.}~\bibnamefont {Eggers}},
  \bibinfo {author} {\bibfnamefont {C.~P.}\ \bibnamefont {Royall}}, \bibinfo
  {author} {\bibfnamefont {S.~R.}\ \bibnamefont {Williams}}, \ and\ \bibinfo
  {author} {\bibfnamefont {H.}~\bibnamefont {Tanaka}},\ }\href@noop {}
  {\bibfield  {journal} {\bibinfo  {journal} {J. Chem. Phys.}\ }\textbf
  {\bibinfo {volume} {138}},\ \bibinfo {pages} {12A535} (\bibinfo {year}
  {2013})}\BibitemShut {NoStop}%
\bibitem [{\citenamefont {Bouchaud}\ and\ \citenamefont
  {Biroli}(2004)}]{Bouchaud-JCP2004}%
  \BibitemOpen
  \bibfield  {author} {\bibinfo {author} {\bibfnamefont {J.-P.}\ \bibnamefont
  {Bouchaud}}\ and\ \bibinfo {author} {\bibfnamefont {G.}~\bibnamefont
  {Biroli}},\ }\href@noop {} {\bibfield  {journal} {\bibinfo  {journal} {J.
  Chem. Phys.}\ }\textbf {\bibinfo {volume} {121}},\ \bibinfo {pages} {7347}
  (\bibinfo {year} {2004})}\BibitemShut {NoStop}%
\bibitem [{\citenamefont {Berthier}\ and\ \citenamefont
  {Kob}(2012)}]{Berthier-static-PRE2012}%
  \BibitemOpen
  \bibfield  {author} {\bibinfo {author} {\bibfnamefont {L.}~\bibnamefont
  {Berthier}}\ and\ \bibinfo {author} {\bibfnamefont {W.}~\bibnamefont {Kob}},\
  }\href@noop {} {\bibfield  {journal} {\bibinfo  {journal} {Phys. Rev. E}\
  }\textbf {\bibinfo {volume} {85}},\ \bibinfo {pages} {011102} (\bibinfo
  {year} {2012})}\BibitemShut {NoStop}%
\bibitem [{\citenamefont {Biroli}\ \emph {et~al.}(2008)\citenamefont {Biroli},
  \citenamefont {Bouchaud}, \citenamefont {Cavagna}, \citenamefont {Grigera},\
  and\ \citenamefont {Verrocchio}}]{Biroli-NatPhys2008}%
  \BibitemOpen
  \bibfield  {author} {\bibinfo {author} {\bibfnamefont {G.}~\bibnamefont
  {Biroli}}, \bibinfo {author} {\bibfnamefont {J.-P.}\ \bibnamefont
  {Bouchaud}}, \bibinfo {author} {\bibfnamefont {A.}~\bibnamefont {Cavagna}},
  \bibinfo {author} {\bibfnamefont {T.}~\bibnamefont {Grigera}}, \ and\
  \bibinfo {author} {\bibfnamefont {P.}~\bibnamefont {Verrocchio}},\
  }\href@noop {} {\bibfield  {journal} {\bibinfo  {journal} {Nat. Phys.}\
  }\textbf {\bibinfo {volume} {4}},\ \bibinfo {pages} {771} (\bibinfo {year}
  {2008})}\BibitemShut {NoStop}%
\bibitem [{\citenamefont {Hocky}\ \emph {et~al.}(2012)\citenamefont {Hocky},
  \citenamefont {Markland},\ and\ \citenamefont {Reichman}}]{Hocky-PRL2012}%
  \BibitemOpen
  \bibfield  {author} {\bibinfo {author} {\bibfnamefont {G.~M.}\ \bibnamefont
  {Hocky}}, \bibinfo {author} {\bibfnamefont {T.~E.}\ \bibnamefont {Markland}},
  \ and\ \bibinfo {author} {\bibfnamefont {D.~R.}\ \bibnamefont {Reichman}},\
  }\href@noop {} {\bibfield  {journal} {\bibinfo  {journal} {Phys. Rev. Let.}\
  }\textbf {\bibinfo {volume} {108}},\ \bibinfo {pages} {225506} (\bibinfo
  {year} {2012})}\BibitemShut {NoStop}%
\bibitem [{\citenamefont {Charbonneau}\ \emph {et~al.}(2012)\citenamefont
  {Charbonneau}, \citenamefont {Charbonneau},\ and\ \citenamefont
  {Tarjus}}]{Charbonneau_Charbonneau_Tarjus_2012}%
  \BibitemOpen
  \bibfield  {author} {\bibinfo {author} {\bibfnamefont {B.}~\bibnamefont
  {Charbonneau}}, \bibinfo {author} {\bibfnamefont {P.}~\bibnamefont
  {Charbonneau}}, \ and\ \bibinfo {author} {\bibfnamefont {G.}~\bibnamefont
  {Tarjus}},\ }\href@noop {} {\bibfield  {journal} {\bibinfo  {journal} {Phys.
  Rev. Lett.}\ }\textbf {\bibinfo {volume} {108}},\ \bibinfo {pages} {035701}
  (\bibinfo {year} {2012})}\BibitemShut {NoStop}%
\bibitem [{\citenamefont {Karmakar}\ \emph {et~al.}(2009)\citenamefont
  {Karmakar}, \citenamefont {Dasgupta},\ and\ \citenamefont
  {Sastry}}]{Karmakar-PNAS2009}%
  \BibitemOpen
  \bibfield  {author} {\bibinfo {author} {\bibfnamefont {S.}~\bibnamefont
  {Karmakar}}, \bibinfo {author} {\bibfnamefont {C.}~\bibnamefont {Dasgupta}},
  \ and\ \bibinfo {author} {\bibfnamefont {S.}~\bibnamefont {Sastry}},\
  }\href@noop {} {\bibfield  {journal} {\bibinfo  {journal} {Proc. Natl. Acad.
  Sci.}\ }\textbf {\bibinfo {volume} {106}},\ \bibinfo {pages} {3675} (\bibinfo
  {year} {2009})}\BibitemShut {NoStop}%
\bibitem [{\citenamefont {Biroli}\ \emph {et~al.}(2013)\citenamefont {Biroli},
  \citenamefont {Karmakar},\ and\ \citenamefont {Procaccia}}]{Biroli-PRL2013}%
  \BibitemOpen
  \bibfield  {author} {\bibinfo {author} {\bibfnamefont {G.}~\bibnamefont
  {Biroli}}, \bibinfo {author} {\bibfnamefont {S.}~\bibnamefont {Karmakar}}, \
  and\ \bibinfo {author} {\bibfnamefont {I.}~\bibnamefont {Procaccia}},\
  }\href@noop {} {\bibfield  {journal} {\bibinfo  {journal} {Phys. Rev. Lett.}\
  }\textbf {\bibinfo {volume} {111}},\ \bibinfo {pages} {165701} (\bibinfo
  {year} {2013})}\BibitemShut {NoStop}%
\bibitem [{\citenamefont {Karmakar}\ \emph {et~al.}(2014)\citenamefont
  {Karmakar}, \citenamefont {Dasgupta},\ and\ \citenamefont
  {Sastry}}]{Karmakar-ARCMP2014}%
  \BibitemOpen
  \bibfield  {author} {\bibinfo {author} {\bibfnamefont {S.}~\bibnamefont
  {Karmakar}}, \bibinfo {author} {\bibfnamefont {C.}~\bibnamefont {Dasgupta}},
  \ and\ \bibinfo {author} {\bibfnamefont {S.}~\bibnamefont {Sastry}},\
  }\href@noop {} {\bibfield  {journal} {\bibinfo  {journal} {Annu. Rev.
  Condens. Matter Phys}\ }\textbf {\bibinfo {volume} {5}},\ \bibinfo {pages}
  {255} (\bibinfo {year} {2014})}\BibitemShut {NoStop}%
\bibitem [{\citenamefont {Ronhovde}\ \emph {et~al.}(2011)\citenamefont
  {Ronhovde}, \citenamefont {Chakrabarty}, \citenamefont {Hu}, \citenamefont
  {Sahu}, \citenamefont {Sahu}, \citenamefont {Kelton}, \citenamefont {Mauro},\
  and\ \citenamefont {Nussinov}}]{Ronhovde-EPJE2011}%
  \BibitemOpen
  \bibfield  {author} {\bibinfo {author} {\bibfnamefont {P.}~\bibnamefont
  {Ronhovde}}, \bibinfo {author} {\bibfnamefont {S.}~\bibnamefont
  {Chakrabarty}}, \bibinfo {author} {\bibfnamefont {D.}~\bibnamefont {Hu}},
  \bibinfo {author} {\bibfnamefont {M.}~\bibnamefont {Sahu}}, \bibinfo {author}
  {\bibfnamefont {K.~K.}\ \bibnamefont {Sahu}}, \bibinfo {author}
  {\bibfnamefont {K.~F.}\ \bibnamefont {Kelton}}, \bibinfo {author}
  {\bibfnamefont {N.~A.}\ \bibnamefont {Mauro}}, \ and\ \bibinfo {author}
  {\bibfnamefont {Z.}~\bibnamefont {Nussinov}},\ }\href@noop {} {\bibfield
  {journal} {\bibinfo  {journal} {Eur. Phys. J. E}\ }\textbf {\bibinfo {volume}
  {34}},\ \bibinfo {pages} {105} (\bibinfo {year} {2011})}\BibitemShut
  {NoStop}%
\bibitem [{\citenamefont {Sausset}\ and\ \citenamefont
  {Levine}(2011)}]{Sausset-PRL2011}%
  \BibitemOpen
  \bibfield  {author} {\bibinfo {author} {\bibfnamefont {F.}~\bibnamefont
  {Sausset}}\ and\ \bibinfo {author} {\bibfnamefont {D.}~\bibnamefont
  {Levine}},\ }\href@noop {} {\bibfield  {journal} {\bibinfo  {journal} {Phys.
  Rev. Lett.}\ }\textbf {\bibinfo {volume} {107}},\ \bibinfo {pages} {045501}
  (\bibinfo {year} {2011})}\BibitemShut {NoStop}%
\bibitem [{\citenamefont {Sammut}\ and\ \citenamefont
  {Webb}(2011)}]{Sammut-Encyclopedia2011}%
  \BibitemOpen
  \bibfield  {author} {\bibinfo {author} {\bibfnamefont {C.}~\bibnamefont
  {Sammut}}\ and\ \bibinfo {author} {\bibfnamefont {G.~I.}\ \bibnamefont
  {Webb}},\ }\href@noop {} {\emph {\bibinfo {title} {Encyclopedia of machine
  learning}}}\ (\bibinfo  {publisher} {Springer-Verlag},\ \bibinfo {year}
  {2011})\BibitemShut {NoStop}%
\bibitem [{\citenamefont {Tanaka}(2012)}]{Tanaka_2012}%
  \BibitemOpen
  \bibfield  {author} {\bibinfo {author} {\bibfnamefont {H.}~\bibnamefont
  {Tanaka}},\ }\href {\doibase 10.1140/epje/i2012-12113-y} {\bibfield
  {journal} {\bibinfo  {journal} {Eur. Phys. J. E}\ }\textbf {\bibinfo {volume}
  {35}},\ \bibinfo {pages} {113} (\bibinfo {year} {2012})}\BibitemShut
  {NoStop}%
\bibitem [{\citenamefont {Langer}(2013)}]{Langer_2013}%
  \BibitemOpen
  \bibfield  {author} {\bibinfo {author} {\bibfnamefont {J.~S.}\ \bibnamefont
  {Langer}},\ }\href {\doibase 10.1103/PhysRevE.88.012122} {\bibfield
  {journal} {\bibinfo  {journal} {Phys. Rev. E}\ }\textbf {\bibinfo {volume}
  {88}},\ \bibinfo {pages} {012122} (\bibinfo {year} {2013})}\BibitemShut
  {NoStop}%
\bibitem [{\citenamefont {Kob}\ and\ \citenamefont
  {Andersen}(1994)}]{Kob-PRL1994}%
  \BibitemOpen
  \bibfield  {author} {\bibinfo {author} {\bibfnamefont {W.}~\bibnamefont
  {Kob}}\ and\ \bibinfo {author} {\bibfnamefont {H.~C.}\ \bibnamefont
  {Andersen}},\ }\href@noop {} {\bibfield  {journal} {\bibinfo  {journal}
  {Phys. Rev. Lett.}\ }\textbf {\bibinfo {volume} {73}},\ \bibinfo {pages}
  {1376} (\bibinfo {year} {1994})}\BibitemShut {NoStop}%
\bibitem [{\citenamefont {Wahnstr{\"o}m}(1991)}]{Wahnstrom-PRA1991}%
  \BibitemOpen
  \bibfield  {author} {\bibinfo {author} {\bibfnamefont {G.}~\bibnamefont
  {Wahnstr{\"o}m}},\ }\href@noop {} {\bibfield  {journal} {\bibinfo  {journal}
  {Physical Review A}\ }\textbf {\bibinfo {volume} {44}},\ \bibinfo {pages}
  {3752} (\bibinfo {year} {1991})}\BibitemShut {NoStop}%
\bibitem [{\citenamefont {Coslovich}(2011)}]{Coslovich-PRE2011}%
  \BibitemOpen
  \bibfield  {author} {\bibinfo {author} {\bibfnamefont {D.}~\bibnamefont
  {Coslovich}},\ }\href {\doibase 10.1103/PhysRevE.83.051505} {\bibfield
  {journal} {\bibinfo  {journal} {Phys. Rev. E}\ }\textbf {\bibinfo {volume}
  {83}},\ \bibinfo {pages} {051505} (\bibinfo {year} {2011})}\BibitemShut
  {NoStop}%
\bibitem [{\citenamefont {Kob}\ \emph {et~al.}(2011)\citenamefont {Kob},
  \citenamefont {Rold{\'a}n-Vargas},\ and\ \citenamefont
  {Berthier}}]{Kob-NatPhys2011}%
  \BibitemOpen
  \bibfield  {author} {\bibinfo {author} {\bibfnamefont {W.}~\bibnamefont
  {Kob}}, \bibinfo {author} {\bibfnamefont {S.}~\bibnamefont
  {Rold{\'a}n-Vargas}}, \ and\ \bibinfo {author} {\bibfnamefont
  {L.}~\bibnamefont {Berthier}},\ }\href@noop {} {\bibfield  {journal}
  {\bibinfo  {journal} {Nature Phys.}\ }\textbf {\bibinfo {volume} {8}},\
  \bibinfo {pages} {164} (\bibinfo {year} {2011})}\BibitemShut {NoStop}%
\bibitem [{Note1()}]{Note1}%
  \BibitemOpen
  \bibinfo {note} {See Supplemental Material [url], which includes Refs. \cite
  {Kim_Saito_2013,OHern_Langer_Liu_Nagel_2002,Stillinger_1976-sup,Lang_Likos_Watzlawek_Lowen_2000-sup,tanemura_geometrical_1977,Biroli-JPCM2007,
  Eaves-PNAS2009-sup,Charbonneau-PRE2010-sup,Charbonneau-JCP2013-sup}.}\BibitemShut
  {Stop}%
\bibitem [{\citenamefont {Schr{\o}der}\ \emph {et~al.}(2000)\citenamefont
  {Schr{\o}der}, \citenamefont {Sastry}, \citenamefont {Dyre},\ and\
  \citenamefont {Glotzer}}]{Schrøder_Sastry_Dyre_Glotzer_2000}%
  \BibitemOpen
  \bibfield  {author} {\bibinfo {author} {\bibfnamefont {T.~B.}\ \bibnamefont
  {Schr{\o}der}}, \bibinfo {author} {\bibfnamefont {S.}~\bibnamefont {Sastry}},
  \bibinfo {author} {\bibfnamefont {J.~C.}\ \bibnamefont {Dyre}}, \ and\
  \bibinfo {author} {\bibfnamefont {S.~C.}\ \bibnamefont {Glotzer}},\ }\href
  {\doibase 10.1063/1.481621} {\bibfield  {journal} {\bibinfo  {journal} {J.
  Chem. Phys.}\ }\textbf {\bibinfo {volume} {112}},\ \bibinfo {pages} {9834–}
  (\bibinfo {year} {2000})}\BibitemShut {NoStop}%
\bibitem [{\citenamefont {Berthier}\ \emph {et~al.}(2012)\citenamefont
  {Berthier}, \citenamefont {Biroli}, \citenamefont {Coslovich}, \citenamefont
  {Kob},\ and\ \citenamefont {Toninelli}}]{Berthier-finite-PRE2012}%
  \BibitemOpen
  \bibfield  {author} {\bibinfo {author} {\bibfnamefont {L.}~\bibnamefont
  {Berthier}}, \bibinfo {author} {\bibfnamefont {G.}~\bibnamefont {Biroli}},
  \bibinfo {author} {\bibfnamefont {D.}~\bibnamefont {Coslovich}}, \bibinfo
  {author} {\bibfnamefont {W.}~\bibnamefont {Kob}}, \ and\ \bibinfo {author}
  {\bibfnamefont {C.}~\bibnamefont {Toninelli}},\ }\href@noop {} {\bibfield
  {journal} {\bibinfo  {journal} {Phys. Rev. E}\ }\textbf {\bibinfo {volume}
  {86}},\ \bibinfo {pages} {031502} (\bibinfo {year} {2012})}\BibitemShut
  {NoStop}%
\bibitem [{\citenamefont {Ikeda}\ and\ \citenamefont
  {Miyazaki}(2011{\natexlab{a}})}]{Ikeda-PRL2011}%
  \BibitemOpen
  \bibfield  {author} {\bibinfo {author} {\bibfnamefont {A.}~\bibnamefont
  {Ikeda}}\ and\ \bibinfo {author} {\bibfnamefont {K.}~\bibnamefont
  {Miyazaki}},\ }\href@noop {} {\bibfield  {journal} {\bibinfo  {journal}
  {Phys. Rev. Lett.}\ }\textbf {\bibinfo {volume} {106}},\ \bibinfo {pages}
  {015701} (\bibinfo {year} {2011}{\natexlab{a}})}\BibitemShut {NoStop}%
\bibitem [{\citenamefont {Ikeda}\ and\ \citenamefont
  {Miyazaki}(2011{\natexlab{b}})}]{Ikeda-JCP2011}%
  \BibitemOpen
  \bibfield  {author} {\bibinfo {author} {\bibfnamefont {A.}~\bibnamefont
  {Ikeda}}\ and\ \bibinfo {author} {\bibfnamefont {K.}~\bibnamefont
  {Miyazaki}},\ }\href@noop {} {\bibfield  {journal} {\bibinfo  {journal} {J.
  Chem. Phys.}\ }\textbf {\bibinfo {volume} {135}},\ \bibinfo {pages} {054901}
  (\bibinfo {year} {2011}{\natexlab{b}})}\BibitemShut {NoStop}%
\bibitem [{Note2()}]{Note2}%
  \BibitemOpen
  \bibinfo {note} {It may appear that the correlation between $n_{\protect \rm
  LPS}$ and dynamics in both the KA and GCM system are superficially similar.
  However, the LPS in the GCM system are simple crystalline motifs that exist
  because of the difficulty of avoiding such particle arrangements in a
  monatomic system. In this sense, we view the correlation of {\protect \em
  both} $n_{\protect \rm LPS}$ and $E_i$ as significantly weaker in the GCM
  system when compared to KA.}\BibitemShut {Stop}%
\bibitem [{\citenamefont {Berthier}\ and\ \citenamefont
  {Jack}(2007)}]{Berthier-PRE2007}%
  \BibitemOpen
  \bibfield  {author} {\bibinfo {author} {\bibfnamefont {L.}~\bibnamefont
  {Berthier}}\ and\ \bibinfo {author} {\bibfnamefont {R.~L.}\ \bibnamefont
  {Jack}},\ }\href@noop {} {\bibfield  {journal} {\bibinfo  {journal} {Phys.
  Rev. E}\ }\textbf {\bibinfo {volume} {76}},\ \bibinfo {pages} {041509}
  (\bibinfo {year} {2007})}\BibitemShut {NoStop}%
\bibitem [{\citenamefont {Franz}\ \emph {et~al.}(2011)\citenamefont {Franz},
  \citenamefont {Parisi}, \citenamefont {Ricci-Tersenghi},\ and\ \citenamefont
  {Rizzo}}]{Franz-EPJE2011}%
  \BibitemOpen
  \bibfield  {author} {\bibinfo {author} {\bibfnamefont {S.}~\bibnamefont
  {Franz}}, \bibinfo {author} {\bibfnamefont {G.}~\bibnamefont {Parisi}},
  \bibinfo {author} {\bibfnamefont {F.}~\bibnamefont {Ricci-Tersenghi}}, \ and\
  \bibinfo {author} {\bibfnamefont {T.}~\bibnamefont {Rizzo}},\ }\href@noop {}
  {\bibfield  {journal} {\bibinfo  {journal} {Eur. Phys. J. E}\ }\textbf
  {\bibinfo {volume} {34}},\ \bibinfo {pages} {1} (\bibinfo {year}
  {2011})}\BibitemShut {NoStop}%
\bibitem [{\citenamefont {Plimpton}(1995)}]{Plimpton-JCP1995}%
  \BibitemOpen
  \bibfield  {author} {\bibinfo {author} {\bibfnamefont {S.}~\bibnamefont
  {Plimpton}},\ }\href {http://lammps.sandia.gov} {\bibfield  {journal}
  {\bibinfo  {journal} {J. Comp. Phys.}\ }\textbf {\bibinfo {volume} {117}},\
  \bibinfo {pages} {1} (\bibinfo {year} {1995})}\BibitemShut {NoStop}%
\bibitem [{\citenamefont {Wilde}\ \emph {et~al.}(2011)\citenamefont {Wilde},
  \citenamefont {Hategan}, \citenamefont {Wozniak}, \citenamefont {Clifford},
  \citenamefont {Katz},\ and\ \citenamefont {Foster}}]{Wilde-Parallel2011}%
  \BibitemOpen
  \bibfield  {author} {\bibinfo {author} {\bibfnamefont {M.}~\bibnamefont
  {Wilde}}, \bibinfo {author} {\bibfnamefont {M.}~\bibnamefont {Hategan}},
  \bibinfo {author} {\bibfnamefont {J.~M.}\ \bibnamefont {Wozniak}}, \bibinfo
  {author} {\bibfnamefont {B.}~\bibnamefont {Clifford}}, \bibinfo {author}
  {\bibfnamefont {D.~S.}\ \bibnamefont {Katz}}, \ and\ \bibinfo {author}
  {\bibfnamefont {I.}~\bibnamefont {Foster}},\ }\href@noop {} {\bibfield
  {journal} {\bibinfo  {journal} {Parallel Comput.}\ }\textbf {\bibinfo
  {volume} {37}},\ \bibinfo {pages} {633 } (\bibinfo {year}
  {2011})}\BibitemShut {NoStop}%
\bibitem [{\citenamefont {Kim}\ and\ \citenamefont
  {Saito}(2013)}]{Kim_Saito_2013}%
  \BibitemOpen
  \bibfield  {author} {\bibinfo {author} {\bibfnamefont {K.}~\bibnamefont
  {Kim}}\ and\ \bibinfo {author} {\bibfnamefont {S.}~\bibnamefont {Saito}},\
  }\href {\doibase doi:10.1063/1.4769256} {\bibfield  {journal} {\bibinfo
  {journal} {J. Chem. Phys.}\ }\textbf {\bibinfo {volume} {138}},\ \bibinfo
  {pages} {12A506} (\bibinfo {year} {2013})}\BibitemShut {NoStop}%
\bibitem [{\citenamefont {O'Hern}\ \emph {et~al.}(2002)\citenamefont {O'Hern},
  \citenamefont {Langer}, \citenamefont {Liu},\ and\ \citenamefont
  {Nagel}}]{OHern_Langer_Liu_Nagel_2002}%
  \BibitemOpen
  \bibfield  {author} {\bibinfo {author} {\bibfnamefont {C.~S.}\ \bibnamefont
  {O'Hern}}, \bibinfo {author} {\bibfnamefont {S.~A.}\ \bibnamefont {Langer}},
  \bibinfo {author} {\bibfnamefont {A.~J.}\ \bibnamefont {Liu}}, \ and\
  \bibinfo {author} {\bibfnamefont {S.~R.}\ \bibnamefont {Nagel}},\ }\href
  {\doibase 10.1103/PhysRevLett.88.075507} {\bibfield  {journal} {\bibinfo
  {journal} {Phys. Rev. Lett.}\ }\textbf {\bibinfo {volume} {88}},\ \bibinfo
  {pages} {075507} (\bibinfo {year} {2002})}\BibitemShut {NoStop}%
\bibitem [{\citenamefont {Stillinger}(1976)}]{Stillinger_1976-sup}%
  \BibitemOpen
  \bibfield  {author} {\bibinfo {author} {\bibfnamefont {F.~H.}\ \bibnamefont
  {Stillinger}},\ }\href {\doibase doi:10.1063/1.432891} {\bibfield  {journal}
  {\bibinfo  {journal} {J. Chem. Phys.}\ }\textbf {\bibinfo {volume} {65}},\
  \bibinfo {pages} {3968–} (\bibinfo {year} {1976})}\BibitemShut {NoStop}%
\bibitem [{\citenamefont {Lang}\ \emph {et~al.}(2000)\citenamefont {Lang},
  \citenamefont {Likos}, \citenamefont {Watzlawek},\ and\ \citenamefont
  {Lowen}}]{Lang_Likos_Watzlawek_Lowen_2000-sup}%
  \BibitemOpen
  \bibfield  {author} {\bibinfo {author} {\bibfnamefont {A.}~\bibnamefont
  {Lang}}, \bibinfo {author} {\bibfnamefont {C.~N.}\ \bibnamefont {Likos}},
  \bibinfo {author} {\bibfnamefont {M.}~\bibnamefont {Watzlawek}}, \ and\
  \bibinfo {author} {\bibfnamefont {H.}~\bibnamefont {Lowen}},\ }\href@noop {}
  {\bibfield  {journal} {\bibinfo  {journal} {J. Phys.: Condens. Matter}\
  }\textbf {\bibinfo {volume} {12}},\ \bibinfo {pages} {5087} (\bibinfo {year}
  {2000})}\BibitemShut {NoStop}%
\bibitem [{\citenamefont {Tanemura}\ \emph {et~al.}(1977)\citenamefont
  {Tanemura}, \citenamefont {Hiwatari}, \citenamefont {Matsuda}, \citenamefont
  {Ogawa}, \citenamefont {Ogita},\ and\ \citenamefont
  {Ueda}}]{tanemura_geometrical_1977}%
  \BibitemOpen
  \bibfield  {author} {\bibinfo {author} {\bibfnamefont {M.}~\bibnamefont
  {Tanemura}}, \bibinfo {author} {\bibfnamefont {Y.}~\bibnamefont {Hiwatari}},
  \bibinfo {author} {\bibfnamefont {H.}~\bibnamefont {Matsuda}}, \bibinfo
  {author} {\bibfnamefont {T.}~\bibnamefont {Ogawa}}, \bibinfo {author}
  {\bibfnamefont {N.}~\bibnamefont {Ogita}}, \ and\ \bibinfo {author}
  {\bibfnamefont {A.}~\bibnamefont {Ueda}},\ }\href@noop {} {\bibfield
  {journal} {\bibinfo  {journal} {Prog. Theor. Phys.}\ }\textbf {\bibinfo
  {volume} {58}},\ \bibinfo {pages} {1079} (\bibinfo {year}
  {1977})}\BibitemShut {NoStop}%
\bibitem [{\citenamefont {Biroli}\ and\ \citenamefont
  {Bouchaud}(2007)}]{Biroli-JPCM2007}%
  \BibitemOpen
  \bibfield  {author} {\bibinfo {author} {\bibfnamefont {G.}~\bibnamefont
  {Biroli}}\ and\ \bibinfo {author} {\bibfnamefont {J.-P.}\ \bibnamefont
  {Bouchaud}},\ }\href@noop {} {\bibfield  {journal} {\bibinfo  {journal} {J.
  Phys: Cond. Mat.}\ }\textbf {\bibinfo {volume} {19}},\ \bibinfo {pages}
  {205101} (\bibinfo {year} {2007})}\BibitemShut {NoStop}%
\bibitem [{\citenamefont {Eaves}\ and\ \citenamefont
  {Reichman}(2009)}]{Eaves-PNAS2009-sup}%
  \BibitemOpen
  \bibfield  {author} {\bibinfo {author} {\bibfnamefont {J.~D.}\ \bibnamefont
  {Eaves}}\ and\ \bibinfo {author} {\bibfnamefont {D.~R.}\ \bibnamefont
  {Reichman}},\ }\href@noop {} {\bibfield  {journal} {\bibinfo  {journal}
  {Proc. Natl. Acad. Sci.}\ }\textbf {\bibinfo {volume} {106}},\ \bibinfo
  {pages} {15171} (\bibinfo {year} {2009})}\BibitemShut {NoStop}%
\bibitem [{\citenamefont {Charbonneau}\ \emph {et~al.}(2010)\citenamefont
  {Charbonneau}, \citenamefont {Ikeda}, \citenamefont {van Meel},\ and\
  \citenamefont {Miyazaki}}]{Charbonneau-PRE2010-sup}%
  \BibitemOpen
  \bibfield  {author} {\bibinfo {author} {\bibfnamefont {P.}~\bibnamefont
  {Charbonneau}}, \bibinfo {author} {\bibfnamefont {A.}~\bibnamefont {Ikeda}},
  \bibinfo {author} {\bibfnamefont {J.~A.}\ \bibnamefont {van Meel}}, \ and\
  \bibinfo {author} {\bibfnamefont {K.}~\bibnamefont {Miyazaki}},\ }\href@noop
  {} {\bibfield  {journal} {\bibinfo  {journal} {Phys. Rev. E}\ }\textbf
  {\bibinfo {volume} {81}},\ \bibinfo {pages} {040501} (\bibinfo {year}
  {2010})}\BibitemShut {NoStop}%
\bibitem [{\citenamefont {Charbonneau}\ \emph {et~al.}(2013)\citenamefont
  {Charbonneau}, \citenamefont {Charbonneau}, \citenamefont {Jin},
  \citenamefont {Parisi},\ and\ \citenamefont
  {Zamponi}}]{Charbonneau-JCP2013-sup}%
  \BibitemOpen
  \bibfield  {author} {\bibinfo {author} {\bibfnamefont {B.}~\bibnamefont
  {Charbonneau}}, \bibinfo {author} {\bibfnamefont {P.}~\bibnamefont
  {Charbonneau}}, \bibinfo {author} {\bibfnamefont {Y.}~\bibnamefont {Jin}},
  \bibinfo {author} {\bibfnamefont {G.}~\bibnamefont {Parisi}}, \ and\ \bibinfo
  {author} {\bibfnamefont {F.}~\bibnamefont {Zamponi}},\ }\href@noop {}
  {\bibfield  {journal} {\bibinfo  {journal} {J. Chem. Phys.}\ }\textbf
  {\bibinfo {volume} {139}},\ \bibinfo {pages} {164502} (\bibinfo {year}
  {2013})}\BibitemShut {NoStop}%
\end{thebibliography}

\begin{thebibliography}{17}%
\makeatletter
\providecommand \@ifxundefined [1]{%
 \@ifx{#1\undefined}
}%
\providecommand \@ifnum [1]{%
 \ifnum #1\expandafter \@firstoftwo
 \else \expandafter \@secondoftwo
 \fi
}%
\providecommand \@ifx [1]{%
 \ifx #1\expandafter \@firstoftwo
 \else \expandafter \@secondoftwo
 \fi
}%
\providecommand \natexlab [1]{#1}%
\providecommand \enquote  [1]{``#1''}%
\providecommand \bibnamefont  [1]{#1}%
\providecommand \bibfnamefont [1]{#1}%
\providecommand \citenamefont [1]{#1}%
\providecommand \href@noop [0]{\@secondoftwo}%
\providecommand \href [0]{\begingroup \@sanitize@url \@href}%
\providecommand \@href[1]{\@@startlink{#1}\@@href}%
\providecommand \@@href[1]{\endgroup#1\@@endlink}%
\providecommand \@sanitize@url [0]{\catcode `\\12\catcode `\$12\catcode
  `\&12\catcode `\#12\catcode `\^12\catcode `\_12\catcode `\%12\relax}%
\providecommand \@@startlink[1]{}%
\providecommand \@@endlink[0]{}%
\providecommand \url  [0]{\begingroup\@sanitize@url \@url }%
\providecommand \@url [1]{\endgroup\@href {#1}{\urlprefix }}%
\providecommand \urlprefix  [0]{URL }%
\providecommand \Eprint [0]{\href }%
\providecommand \doibase [0]{http://dx.doi.org/}%
\providecommand \selectlanguage [0]{\@gobble}%
\providecommand \bibinfo  [0]{\@secondoftwo}%
\providecommand \bibfield  [0]{\@secondoftwo}%
\providecommand \translation [1]{[#1]}%
\providecommand \BibitemOpen [0]{}%
\providecommand \bibitemStop [0]{}%
\providecommand \bibitemNoStop [0]{.\EOS\space}%
\providecommand \EOS [0]{\spacefactor3000\relax}%
\providecommand \BibitemShut  [1]{\csname bibitem#1\endcsname}%
\let\auto@bib@innerbib\@empty
\bibitem [{\citenamefont {Kim}\ and\ \citenamefont
  {Saito}(2013)}]{Kim_Saito_2013-sup}%
  \BibitemOpen
  \bibfield  {author} {\bibinfo {author} {\bibfnamefont {K.}~\bibnamefont
  {Kim}}\ and\ \bibinfo {author} {\bibfnamefont {S.}~\bibnamefont {Saito}},\
  }\href {\doibase doi:10.1063/1.4769256} {\bibfield  {journal} {\bibinfo
  {journal} {J. Chem. Phys.}\ }\textbf {\bibinfo {volume} {138}},\ \bibinfo
  {pages} {12A506} (\bibinfo {year} {2013})}\BibitemShut {NoStop}%
\bibitem [{\citenamefont {Kob}\ and\ \citenamefont
  {Andersen}(1994)}]{Kob-PRL1994-sup}%
  \BibitemOpen
  \bibfield  {author} {\bibinfo {author} {\bibfnamefont {W.}~\bibnamefont
  {Kob}}\ and\ \bibinfo {author} {\bibfnamefont {H.}~\bibnamefont {Andersen}},\
  }\href@noop {} {\bibfield  {journal} {\bibinfo  {journal} {Phys. Rev. Lett.}\
  }\textbf {\bibinfo {volume} {73}},\ \bibinfo {pages} {1376} (\bibinfo {year}
  {1994})}\BibitemShut {NoStop}%
\bibitem [{\citenamefont {Kob}\ \emph {et~al.}(2011)\citenamefont {Kob},
  \citenamefont {Rold{\'a}n-Vargas},\ and\ \citenamefont
  {Berthier}}]{Kob-NatPhys2011-sup}%
  \BibitemOpen
  \bibfield  {author} {\bibinfo {author} {\bibfnamefont {W.}~\bibnamefont
  {Kob}}, \bibinfo {author} {\bibfnamefont {S.}~\bibnamefont
  {Rold{\'a}n-Vargas}}, \ and\ \bibinfo {author} {\bibfnamefont
  {L.}~\bibnamefont {Berthier}},\ }\href@noop {} {\bibfield  {journal}
  {\bibinfo  {journal} {Nature Phys.}\ }\textbf {\bibinfo {volume} {8}},\
  \bibinfo {pages} {164} (\bibinfo {year} {2011})}\BibitemShut {NoStop}%
\bibitem [{\citenamefont {Ikeda}\ and\ \citenamefont
  {Miyazaki}(2011{\natexlab{a}})}]{Ikeda-JCP2011-sup}%
  \BibitemOpen
  \bibfield  {author} {\bibinfo {author} {\bibfnamefont {A.}~\bibnamefont
  {Ikeda}}\ and\ \bibinfo {author} {\bibfnamefont {K.}~\bibnamefont
  {Miyazaki}},\ }\href@noop {} {\bibfield  {journal} {\bibinfo  {journal} {J.
  Chem. Phys.}\ }\textbf {\bibinfo {volume} {135}},\ \bibinfo {pages} {054901}
  (\bibinfo {year} {2011}{\natexlab{a}})}\BibitemShut {NoStop}%
\bibitem [{\citenamefont {Berthier}\ and\ \citenamefont
  {Jack}(2007)}]{Berthier-PRE2007-sup}%
  \BibitemOpen
  \bibfield  {author} {\bibinfo {author} {\bibfnamefont {L.}~\bibnamefont
  {Berthier}}\ and\ \bibinfo {author} {\bibfnamefont {R.~L.}\ \bibnamefont
  {Jack}},\ }\href@noop {} {\bibfield  {journal} {\bibinfo  {journal} {Phys.
  Rev. E}\ }\textbf {\bibinfo {volume} {76}},\ \bibinfo {pages} {041509}
  (\bibinfo {year} {2007})}\BibitemShut {NoStop}%
\bibitem [{\citenamefont {Wahnstr{\"o}m}(1991)}]{Wahnstrom-PRA1991-sup}%
  \BibitemOpen
  \bibfield  {author} {\bibinfo {author} {\bibfnamefont {G.}~\bibnamefont
  {Wahnstr{\"o}m}},\ }\href@noop {} {\bibfield  {journal} {\bibinfo  {journal}
  {Phys. Rev. A}\ }\textbf {\bibinfo {volume} {44}},\ \bibinfo {pages} {3752}
  (\bibinfo {year} {1991})}\BibitemShut {NoStop}%
\bibitem [{\citenamefont {O’Hern}\ \emph {et~al.}(2002)\citenamefont
  {O’Hern}, \citenamefont {Langer}, \citenamefont {Liu},\ and\ \citenamefont
  {Nagel}}]{OHern_Langer_Liu_Nagel_2002-sup}%
  \BibitemOpen
  \bibfield  {author} {\bibinfo {author} {\bibfnamefont {C.~S.}\ \bibnamefont
  {O’Hern}}, \bibinfo {author} {\bibfnamefont {S.~A.}\ \bibnamefont
  {Langer}}, \bibinfo {author} {\bibfnamefont {A.~J.}\ \bibnamefont {Liu}}, \
  and\ \bibinfo {author} {\bibfnamefont {S.~R.}\ \bibnamefont {Nagel}},\ }\href
  {\doibase 10.1103/PhysRevLett.88.075507} {\bibfield  {journal} {\bibinfo
  {journal} {Phys. Rev. Lett.}\ }\textbf {\bibinfo {volume} {88}},\ \bibinfo
  {pages} {075507} (\bibinfo {year} {2002})}\BibitemShut {NoStop}%
\bibitem [{\citenamefont {Stillinger}(1976)}]{Stillinger_1976-sup}%
  \BibitemOpen
  \bibfield  {author} {\bibinfo {author} {\bibfnamefont {F.~H.}\ \bibnamefont
  {Stillinger}},\ }\href {\doibase doi:10.1063/1.432891} {\bibfield  {journal}
  {\bibinfo  {journal} {J. Chem. Phys.}\ }\textbf {\bibinfo {volume} {65}},\
  \bibinfo {pages} {3968–3974} (\bibinfo {year} {1976})}\BibitemShut
  {NoStop}%
\bibitem [{\citenamefont {Lang}\ \emph {et~al.}(2000)\citenamefont {Lang},
  \citenamefont {Likos}, \citenamefont {Watzlawek},\ and\ \citenamefont
  {Lowen}}]{Lang_Likos_Watzlawek_Lowen_2000-sup}%
  \BibitemOpen
  \bibfield  {author} {\bibinfo {author} {\bibfnamefont {A.}~\bibnamefont
  {Lang}}, \bibinfo {author} {\bibfnamefont {C.~N.}\ \bibnamefont {Likos}},
  \bibinfo {author} {\bibfnamefont {M.}~\bibnamefont {Watzlawek}}, \ and\
  \bibinfo {author} {\bibfnamefont {H.}~\bibnamefont {Lowen}},\ }\href@noop {}
  {\bibfield  {journal} {\bibinfo  {journal} {J. Phys.: Condens. Matter}\
  }\textbf {\bibinfo {volume} {12}},\ \bibinfo {pages} {5087–5108} (\bibinfo
  {year} {2000})}\BibitemShut {NoStop}%
\bibitem [{\citenamefont {Plimpton}(1995)}]{Plimpton-JCP1995-sup}%
  \BibitemOpen
  \bibfield  {author} {\bibinfo {author} {\bibfnamefont {S.}~\bibnamefont
  {Plimpton}},\ }\href {http://lammps.sandia.gov} {\bibfield  {journal}
  {\bibinfo  {journal} {J. Comp. Phys.}\ }\textbf {\bibinfo {volume} {117}},\
  \bibinfo {pages} {1} (\bibinfo {year} {1995})}\BibitemShut {NoStop}%
\bibitem [{\citenamefont {Coslovich}(2011)}]{Coslovich-PRE2011-sup}%
  \BibitemOpen
  \bibfield  {author} {\bibinfo {author} {\bibfnamefont {D.}~\bibnamefont
  {Coslovich}},\ }\href {\doibase 10.1103/PhysRevE.83.051505} {\bibfield
  {journal} {\bibinfo  {journal} {Phys. Rev. E}\ }\textbf {\bibinfo {volume}
  {83}},\ \bibinfo {pages} {051505} (\bibinfo {year} {2011})}\BibitemShut
  {NoStop}%
\bibitem [{\citenamefont {Tanemura}\ \emph {et~al.}(1977)\citenamefont
  {Tanemura}, \citenamefont {Hiwatari}, \citenamefont {Matsuda}, \citenamefont
  {Ogawa}, \citenamefont {Ogita},\ and\ \citenamefont
  {Ueda}}]{tanemura_geometrical_1977-sup}%
  \BibitemOpen
  \bibfield  {author} {\bibinfo {author} {\bibfnamefont {M.}~\bibnamefont
  {Tanemura}}, \bibinfo {author} {\bibfnamefont {Y.}~\bibnamefont {Hiwatari}},
  \bibinfo {author} {\bibfnamefont {H.}~\bibnamefont {Matsuda}}, \bibinfo
  {author} {\bibfnamefont {T.}~\bibnamefont {Ogawa}}, \bibinfo {author}
  {\bibfnamefont {N.}~\bibnamefont {Ogita}}, \ and\ \bibinfo {author}
  {\bibfnamefont {A.}~\bibnamefont {Ueda}},\ }\href@noop {} {\bibfield
  {journal} {\bibinfo  {journal} {Prog. Theor. Phys.}\ }\textbf {\bibinfo
  {volume} {58}},\ \bibinfo {pages} {1079} (\bibinfo {year}
  {1977})}\BibitemShut {NoStop}%
\bibitem [{\citenamefont {Biroli}\ and\ \citenamefont
  {Bouchaud}(2007)}]{Biroli-JPCM2007-sup}%
  \BibitemOpen
  \bibfield  {author} {\bibinfo {author} {\bibfnamefont {G.}~\bibnamefont
  {Biroli}}\ and\ \bibinfo {author} {\bibfnamefont {J.-P.}\ \bibnamefont
  {Bouchaud}},\ }\href@noop {} {\bibfield  {journal} {\bibinfo  {journal} {J.
  Phys: Cond. Mat.}\ }\textbf {\bibinfo {volume} {19}},\ \bibinfo {pages}
  {205101} (\bibinfo {year} {2007})}\BibitemShut {NoStop}%
\bibitem [{\citenamefont {Eaves}\ and\ \citenamefont
  {Reichman}(2009)}]{Eaves-PNAS2009-sup}%
  \BibitemOpen
  \bibfield  {author} {\bibinfo {author} {\bibfnamefont {J.~D.}\ \bibnamefont
  {Eaves}}\ and\ \bibinfo {author} {\bibfnamefont {D.~R.}\ \bibnamefont
  {Reichman}},\ }\href@noop {} {\bibfield  {journal} {\bibinfo  {journal}
  {Proc. Natl. Acad. Sci.}\ }\textbf {\bibinfo {volume} {106}},\ \bibinfo
  {pages} {15171} (\bibinfo {year} {2009})}\BibitemShut {NoStop}%
\bibitem [{\citenamefont {Charbonneau}\ \emph {et~al.}(2010)\citenamefont
  {Charbonneau}, \citenamefont {Ikeda}, \citenamefont {van Meel},\ and\
  \citenamefont {Miyazaki}}]{Charbonneau-PRE2010-sup}%
  \BibitemOpen
  \bibfield  {author} {\bibinfo {author} {\bibfnamefont {P.}~\bibnamefont
  {Charbonneau}}, \bibinfo {author} {\bibfnamefont {A.}~\bibnamefont {Ikeda}},
  \bibinfo {author} {\bibfnamefont {J.}~\bibnamefont {van Meel}}, \ and\
  \bibinfo {author} {\bibfnamefont {K.}~\bibnamefont {Miyazaki}},\ }\href@noop
  {} {\bibfield  {journal} {\bibinfo  {journal} {Phys. Rev. E}\ }\textbf
  {\bibinfo {volume} {81}},\ \bibinfo {pages} {040501} (\bibinfo {year}
  {2010})}\BibitemShut {NoStop}%
\bibitem [{\citenamefont {Ikeda}\ and\ \citenamefont
  {Miyazaki}(2011{\natexlab{b}})}]{Ikeda-PRL2011-sup}%
  \BibitemOpen
  \bibfield  {author} {\bibinfo {author} {\bibfnamefont {A.}~\bibnamefont
  {Ikeda}}\ and\ \bibinfo {author} {\bibfnamefont {K.}~\bibnamefont
  {Miyazaki}},\ }\href@noop {} {\bibfield  {journal} {\bibinfo  {journal}
  {Phys. Rev. Lett.}\ }\textbf {\bibinfo {volume} {106}},\ \bibinfo {pages}
  {015701} (\bibinfo {year} {2011}{\natexlab{b}})}\BibitemShut {NoStop}%
\bibitem [{\citenamefont {Charbonneau}\ \emph {et~al.}(2013)\citenamefont
  {Charbonneau}, \citenamefont {Charbonneau}, \citenamefont {Jin},
  \citenamefont {Parisi},\ and\ \citenamefont
  {Zamponi}}]{Charbonneau-JCP2013-sup}%
  \BibitemOpen
  \bibfield  {author} {\bibinfo {author} {\bibfnamefont {B.}~\bibnamefont
  {Charbonneau}}, \bibinfo {author} {\bibfnamefont {P.}~\bibnamefont
  {Charbonneau}}, \bibinfo {author} {\bibfnamefont {Y.}~\bibnamefont {Jin}},
  \bibinfo {author} {\bibfnamefont {G.}~\bibnamefont {Parisi}}, \ and\ \bibinfo
  {author} {\bibfnamefont {F.}~\bibnamefont {Zamponi}},\ }\href@noop {}
  {\bibfield  {journal} {\bibinfo  {journal} {J. Chem. Phys.}\ }\textbf
  {\bibinfo {volume} {139}},\ \bibinfo {pages} {164502} (\bibinfo {year}
  {2013})}\BibitemShut {NoStop}%
\end{thebibliography}
\end{document}